\documentclass[epj]{svjour}
\usepackage{graphics,amsmath,amssymb}
\def\d{{\rm d}}

\def\e{{\rm e}}
\def\p{{\rm p}}
\def\i{{\rm i}}
\def\u{{\rm u}}

\def\gs{\gamma_S}
\def\ssb{$\langle {\rm s}\bar{\rm s} \rangle$} 
\def\Oav{{\langle O \rangle}}
\def\Nj{{\{N_j\}}}

\def\Hj{{\{H_j\}}}
\def\Omj{\Omega_{\{N_j\}}}
\def\Pij{\Pi_{\{N_j\}}}

\def\hpartj{\{h_{n_j}\}}
\def\ri{\{r_i\}}
\def\Qz{{\bf Q}}

\def\qj{{\bf q}_j}
\def\delq{{\delta_{\Qz,\sum_j N_j \qj}}}

\def\phivs{\mbox{\boldmath${\scriptstyle{\phi}}$}}
\def\pivs{\mbox{\boldmath${\scriptstyle{\pi}}$}}

\def\ee{e$^+$e$^-$}
\def\ppb{${\rm p}\bar{\rm p}\;$}
\def\eecc{${\rm e}^+{\rm e}^- \rightarrow {\rm c}\bar{\rm c}\;$}
\def\eebb{${\rm e}^+{\rm e}^- \rightarrow {\rm b}\bar{\rm b}\;$}

\begin{document}
\title{Statistical hadronization and hadronic microcanonical ensemble II}
\author{F. Becattini\inst{1} \and L. Ferroni\inst{1}
}                     % Do not remove
\institute{Universit\`a di Firenze and INFN Sezione di Firenze}
\abstract{
We present a Monte-Carlo calculation of the microcanonical ensemble 
of the of the ideal hadron-resonance gas including all known states 
up to a mass of about 1.8 GeV and full quantum statistics. The microcanonical 
average multiplicities of the various hadron species are found to converge 
to the canonical ones for moderately low values of the total energy, 
around 8 GeV, thus bearing out previous analyses of hadronic
multiplicities in the canonical ensemble. The main numerical computing 
method is an importance sampling Monte-Carlo algorithm using the product
of Poisson distributions to generate multi-hadronic channels. It is shown 
that the use of this multi-Poisson distribution allows an efficient and 
fast computation of averages, which can be further improved in the limit of 
very large clusters. We have also studied the fitness of a previously 
proposed computing method, based on the Metropolis Monte-Carlo algorithm, 
for event generation in the statistical hadronization model. We find that 
the use of the multi-Poisson distribution as proposal matrix dramatically 
improves the computation performance. However, due to the correlation of
subsequent samples, this method proves to be generally less robust and 
effective than the importance sampling method.
\PACS{
      {PACS-key}{discribing text of that key}   \and
      {PACS-key}{discribing text of that key}
     } % end of PACS codes
} %end of abstract
\maketitle
%   
%***********************************************************************
\section{Introduction}
%***********************************************************************

The basic assumption of the statistical hadronization model (SHM) is that, 
as a consequence of a dynamical QCD-driven process, the final stage of a 
high energy collision gives rise to the formation a set of colourless massive 
extended objects, defined as {\em clusters} (or {\em fireballs}). They are 
assumed to produce hadrons in a purely statistical manner, namely all 
multihadronic states within the cluster volume and compatible with 
its quantum numbers are equally likely. The set of equiprobable states with 
fixed four-momentum and internal charges (abelian or not) is what is usually 
called the 
microcanonical ensemble \footnote{The full microcanonical ensemble should
in principle include angular momentum and parity conservation. In this work,
as well as in previous literature, angular momentum and parity are disregarded
and the resulting set of states are still defined as microcanonical ensemble.}. 
Interactions between stable hadrons are mostly taken into account by the 
inclusion of all resonances as independent states \cite{bdm}, which is the 
reason of the usual expression, in the SHM, of {\em ideal hadron-resonance gas} 
\cite{hage}. 

The special motivations for a detailed study of the microcanonical ensemble 
of the ideal hadron-resonance gas have been extensively discussed in the first 
paper \cite{beca1}. They can be shortly summarized as:
\begin{enumerate}
\item{} The need of a tool to hadronize final-state clusters in particle 
and heavy ion collisions in the statistical hadronization model and test
observables which cannot be calculated analytically.
\item{} The assessment of the validity of previous calculations in the 
canonical ensemble, especially in the analysis of average multiplicities
in high energy elementary collisions \cite{beca,becaheinz,becagp}.
\end{enumerate}

The first point is quite clear: if we could compute microcanonical averages
numerically in a practical way, we would be able to make predictions on many 
observables within the basic framework of the statistical model without 
invoking further assumptions or approximations which are needed to obtain analytical 
expressions. Furtermore, the availability of a Monte-Carlo integration algorithm  
opens the way to event generators with hadronization stage modelled by SHM. 
The second point is somehow related to the first. Indeed, in previous work 
within SHM, specific assumptions were invoked in order to allow the use of the 
canonical ensemble, which is far easier to handle with regard to both analytical 
and numerical calculations. It is then necessary to verify whether it was sound
to calculate multiplicities in the canonical ensemble by comparing them to those 
in the microcanonical ensemble with the same values of total mass and volume.

The microcanonical formalism for the hadron gas and the relation between 
microcanonical and canonical ensembles have been developed in the first paper
\cite{beca1}. In this paper we will focus on the numerical analysis; we 
will discuss the main computational issues and we will describe two 
algorithms to sample the microcanonical hadron gas phase space. As has been 
mentioned above, we will inspect the differences between microcanonical and 
canonical averages. In this first study, we will only deal with observables 
pertaining to particle multiplicities, leaving the analysis of momentum 
spectra to further works.

The microcanonical ensemble of the hadron gas has been investigated numerically
by K. Werner and J. Aichelin \cite{weai} by using a Monte-Carlo method based on
Metropolis algorithm. Results on specific observables have recently been published
\cite{liu} in a hadron gas model including fundamental multiplets (two meson
nonets and baryon octet plus decaplet). In comparison with these previous works,
we will show calculations for the full hadron gas including all known species 
(more than 250) up to a mass of about 1.8 GeV and, particularly, we will introduce 
a new updating rule for Metropolis algorithm leading to a dramatic improvement 
of its performance in terms of computing time. Moreover, we propose a different 
Monte-Carlo computing method, based on the importance sampling of multi-hadronic 
channel space, which proves to be more effective than Metropolis algorithm for the 
calculation of averages.

The paper is organized as follows: the basic microcanonical formalism is 
summarized in Sect.~2; the numerical method to compute the phase space  
volumes for a given multi-hadronic channel is discussed in Sect. 3; in Sect.~4
we describe the importance sampling method which is well suited to compute
averages in the microcanonical ensemble of the ideal hadron-resonance gas; 
in Sect.~5 we show the comparison between microcanonical and canonical 
averages; in Sect.~6 and 7 the Metropolis algorithm is studied in detail; 
finally, conclusions are summarized in Sect.~8.

%***********************************************************************   
\section{Microcanonical partition function}
%***********************************************************************

In principle, any average on a given statistical mechanics ensemble could be 
calculated from the partition function. The {\em microcanonical partition 
function} of the hadron gas is best defined \cite{beca1} as the sum over all 
multihadronic states localized within the cluster $| h_V \rangle$ constrained 
with four-momentum and abelian (i.e. additive) charge conservation:
\begin{equation}\label{micro1}
 \Omega = \sum_{h_V} \langle h_V | \delta^4 (P- P_{\rm op}) 
 \delta_{\Qz,\Qz_{\rm op}} | h_V \rangle
\end{equation}
where $\Qz = (Q_1,\ldots,Q_M)$ is a vector of $M$ integer abelian charges 
(electric, baryon number, strangeness etc.), $P$ the four-momentum of the 
cluster and $P_{\rm op}$, $\Qz_{\rm op}$ the relevant operators. Provided that 
relativistic quantum field effects are neglected and the volume of the cluster 
is large enough to allow the approximation of finite-volume Fourier integrals 
with Dirac deltas, it can be proved \cite{beca1} that $\Omega$ can be written 
as a multiple integral: 
\begin{eqnarray}\label{micro2}
\!\!\!\!\!\!\!\!\!\!\! && \Omega= \frac{1}{(2\pi)^{4+M}} \int \d^4 y \;\e^{\i P \cdot y} 
\int_{-\pivs}^{+\pivs} \d^M \phi \; \e^{\i \Qz \cdot \phivs} \nonumber\\
\!\!\!\!\!\!\!\!\!\!\! && \times \exp\Big[\sum_j \frac{(2J_j+1)V}{(2\pi)^3} \!\!\! 
 \int \d^3\p \; \log (1\pm \e^{-\i p_j \cdot y -\i \qj \cdot \phivs})^{\pm 1}\Big]  
\end{eqnarray}
where $\qj$ is the vector of the abelian charges for the $j^{\rm th}$ hadron 
species, $J_j$ its spin, $V$ the volume of the cluster; the upper sign applies
to fermions, the lower to bosons. The integral (\ref{micro2}) is more 
easily calculable in the rest frame of the cluster where $P= (M,{\bf 0})$. 
Unfortunately, an analytical solution with no charge constraint (the so-called 
{\em grand-microcanonical partition function}) is known only in two limiting 
cases: non-relativistic and ultra-relativistic (i.e. with all particle masses set 
to zero). The full relativistic case has been attacked with several
kinds of expansions \cite{cerhag1} but none of them proved to be fully satisfactory
as the achieved accuracy in the estimation of different kinds of averages
could vary from some percent to a factor 10. Therefore, a numerical integration of
Eq. (\ref{micro2}) is needed. The most suitable method is to decompose $\Omega$
into the sum of the phase space volumes with fixed particle multiplicities for each
species:
\begin{equation}\label{microdec}  
  \Omega = \sum_{\Nj} \Omega_{\Nj} \delq
\end{equation}
$\Nj$ being a vector of $K$ integer numbers $(N_1,\ldots,N_K)$, i.e. the multiplicities
of all of the $K$ hadronic species. The set of integers $\Nj$ is also defined as 
a {\em channel} because it characterizes a specific decay channel of the cluster. 
The general phase space volume $\Omj$ for the channel $\Nj$ obtained from can be 
written as a cluster decomposition. Let $j$ be the integer index running over all 
hadron species, and $\hpartj$ a {\em partition} (relevant to the species $j$) of 
$N_j$ in the multiplicity representation, i.e. a set of integers such that 
$N_j = \sum_{n_j=1}^{N_j} n_j h_{n_j}$; let $H_j = \sum_{n_j=1}^{N_j} h_{n_j}$ 
and let $c_{l_j}$ be the cyclic permutations of the first $n_{l_j}$ integers
determined by the partition, with $\sum_{l_j=1}^{H_j} n_{l_j} = N_j$. Provided 
that relativistic quantum field effects are neglected (which is possible if the 
cluster linear size is greater than pion Compton wavelength, 1.4 fm), the phase 
space volume for fixed multiplicities reads \cite{beca1}:
\begin{eqnarray}\label{clusexp0}
\!\!\!\!\!\!\!\!\Omega_{\Nj} &=& \int \d^3 \p_1 \ldots \d^3 \p_N \; 
\delta^4 (P-P_f)  \nonumber \\
\!\!\!\!\!\!\!\! &\times& \prod_j \sum_{\hpartj} \frac{(\mp 1)^{N_j + H_j} 
 (2J_j + 1)^{H_j}}
 {\prod_{n_j=1}^{N_j} n_j^{h_{n_j}} h_{n_j}!} \prod_{l_j=1}^{H_j} F_{n_{l_j}}
\end{eqnarray}
where $P_f$ is the sum of all particle four-momenta. The factors $F_{n_{l_j}}$ in the 
equation above are Fourier integrals over the cluster region with proper volume $V$:
\begin{equation}\label{fint}
 F_{n_{l_j}} = \prod_{i_{l_j}=1}^{n_{l_j}} \frac{1}{(2\pi)^3} \int_V \d^3 {\rm x} \; 
 \e^{\i {\bf x \cdot}({\bf p}_{i_{l_j}} - {\bf p}_{c_{l_j}(i_{l_j})})}
\end{equation}
and the ${\bf p}$'s are the particle momenta. For sufficiently large $V$, the integral 
in Eq. (\ref{fint}) tends to a product of Dirac delta distribution and, if we use this 
limit in Eq. (\ref{clusexp0}), we arrive at this expression of $\Omj$:
\begin{eqnarray}\label{clusexp}
\!\!\!\!\!\!\!\!\Omega_{\Nj} &=& 
\Bigg[ \prod_j \sum_{\hpartj} (\mp 1)^{N_j + H_j} \frac{1}{\prod_{n_j=1}^{N_j}
 n_j^{4h_{n_j}} h_{n_j}!} \nonumber \\
\!\!\!\!\!\!\!\! && \Big[ \prod_{l_j=1}^{H_j} \frac{V (2J_j+1)}{(2\pi)^3} 
  \int \d^3 \p'_{l_j}\Big]\Bigg] \delta^4 (P - \!\!\! \sum_{j,l_j=1}^{H_j} 
 \!\! p'_{l_j} )  
\end{eqnarray}
where, for a set of partitions ${\{h_{n_1}\}}, \ldots, {\{h_{n_K}\}}$ for each
of the hadron species, the four-momenta $p'_{l_j}$ are those of lumps of particles 
of the same species $j$ ($H_j$ in number) with mass $n_j m_j$ and spin $J_j$. 

For sufficiently large volumes, the leading term in Eq. (\ref{clusexp}), henceforth 
defined as $\Omj^c$, is that with the maximal power of $V$, i.e. with $H_j = N_j \; 
\forall j$. This term corresponds to the partitions ${\hpartj}=(N_j,0,\ldots) \; 
\forall j$, namely to one particle per lump, and reads:
\begin{eqnarray}\label{boltz}
\!\!\!\!\!\!\!\!\!  && \Omj^c = \nonumber \\
\!\!\!\!\!\!\!\!\!  && = \prod_j \frac{V^{N_j}(2J_j+1)^{N_j}}{(2\pi)^{3N_j} N_j!}  
  \int \d^3 \p_1 \ldots \d^3 \p_N \; \delta^4 (P-\sum_{i=1}^N p_i)  
\end{eqnarray}     
where $N = \sum_j N_j$. This is indeed the phase space volume in the classical 
Boltzmann statistics. Therefore, the terms beyond the leading one (\ref{boltz}) 
in the expansions (\ref{clusexp0}) and (\ref{clusexp}) account for the quantum 
statistics effects. The very same expression (\ref{boltz}) holds as the leading
term of the more general cluster decomposition (\ref{clusexp0}). In fact, the
cyclic permutations $c_{l_j}$ corresponding to the partition ${\hpartj}=(N_j,0,\ldots)$
are identities, thus implying, according to Eq. (\ref{fint}):
\begin{equation}
  \prod_{l_j=1}^{N_j} F_{n_{l_j}} = \frac{V^{N_j}}{(2\pi)^{3N_j}}
\end{equation}
In the present work, we have used the approximated expression (\ref{clusexp})
of Eq. (\ref{clusexp0}) to evaluate the phase space volume for fixed multiplicities.
For the considered cluster masses and volumes (see later on) this approximation 
is satisfactory for most observables, taking into account that the leading term 
(\ref{boltz}) of the cluster decomposition is the same in both Eq. (\ref{clusexp0})
and (\ref{clusexp}) and that subleading terms give at most a 10\% correction to the 
leading term.  
  
A nice feature of the cluster decomposition (\ref{clusexp}) is that every term 
of the expansion is an integral just like the classical phase space volume 
(\ref{boltz}) with lumps replacing particles. Specifically, the Eq. (\ref{clusexp}) 
can be rewritten as:
\begin{equation}\label{clusexp2}
 \Omega_{\Nj} = 
 \sum_{\{h_{n_1}\},\ldots,\{h_{n_K}\}} \Bigg[ \prod_j \frac{(\mp 1)^{N_j + H_j}} 
 {\prod_{n_j=1}^{N_j} n_j^{4h_{n_j}}} \Bigg] \Omega_{\Hj}^c  
\end{equation}
Note that the factor $\prod_{n_j=1}^{N_j} h_{n_j}!$ has been absorbed in 
$\Omega_{\Hj}^c$ as it takes into account the identity of the lumps. This form
of the cluster decomposition shows that in actual numerical calculations all
of the terms can be computed with the same routine. 

As has been mentioned in the introduction, in this paper we are mainly 
interested in the calculation of quantities relevant to particle multiplicities 
and not to their momenta, namely their kinematical state. The average of an 
observable $O$ depending on particle multiplicities in the microcanonical 
ensemble can then be written as:
\begin{equation}\label{obs}
  \langle O \rangle = \frac{\sum_{\Nj} O(\Nj) \Omj \delq}
  {\sum_{\Nj} \Omj \delq}
\end{equation}
Altogether, what we need to calculate in order to evaluate an average~(\ref{obs})
are integrals like~(\ref{clusexp}) and~(\ref{boltz}). The description of a 
suitable numerical technique to do that is the subject of next section. 
   
%***********************************************************************   
\section{Numerical calculation of the phase space volume}
%***********************************************************************

In order to calculate efficiently and quickly the phase space volume for fixed 
multiplicities $\Omj$ in Eq. (\ref{clusexp}), we have adopted a Monte-Carlo 
integration method proposed by Cerulus and Hagedorn in the 60's \cite{cerhag1,cerhag2} 
and later employed by Werner and Aichelin \cite{weai}. The method is 
described in detail in ref.~\cite{weai}. As we made only slight modifications,
we just sketch it here; a more detailed description is given in Appendix A. 
As has already been mentioned, every term in the cluster decomposition 
(\ref{clusexp}) is an integral of the kind (\ref{boltz}) and can be calculated 
by the same numerical method. 

The calculation is carried out in the cluster's rest frame, where $P=(M,{\bf 0})$ 
and $V$ is the proper volume. The integral~(\ref{boltz}) is first written 
as the product:
\begin{equation}   
 \Omj^c = \frac{V^N T^{3N-4}}{(2\pi)^{3N}}\prod_j \frac{(2J_j+1)^{N_j}}{N_j!} \;
 \Phi (M,m_1,\ldots,m_N)
\end{equation}
where $T = M - \sum_{i=1}^N m_i$ is the available total kinetic energy for the $N$ 
particles (or lumps), and $\Phi$ is an adimensional kinematic integral:
\begin{equation}\label{Phi}
 \Phi = \frac{1}{T^{3N-4}}
 \int \d^3 \p_1 \ldots \d^3 \p_N \; \delta^4(P-\sum_{i=1}^N p_i) 
\end{equation}
After a sequence of variable changes, the function $\Phi$ is rewritten as an
integral of an adimensional function $\Upsilon$ of $N-1$ variables $r_i 
\in [0,1]$ (see Appendix A):
\begin{equation}\label{Upsilon}
 \Phi = \int_0^1 \d r_1 \ldots \int_0^1 \d r_{N-1} \; \Upsilon (r_1,\ldots,r_{N-1})
\end{equation}
which can be estimated through Monte-Carlo integration as:
\begin{equation}\label{mcphi}
 \Phi \doteq \frac{1}{N_S} \sum_{k=1}^{N_S} \Upsilon (r_1^{(k)},\ldots,r_{N-1}^{(k)})
\end{equation}
where $r_i^{(k)}$ are random numbers uniformely distributed in the interval $[0,1]$
and $N_S$ is the number of samples.

In order to calculate the full cluster decomposition of the phase space volume, 
in Eq. (\ref{clusexp2}), the above calculation is repeated for all of the terms.
To reduce the number of calls to the random number generation subroutine, one can 
take advantage of rewriting Eq. (\ref{clusexp2}) as:
\begin{eqnarray}\label{clusexp3}
 \Omega_{\Nj} = && \sum_{\{h_{n_1}\},\ldots,\{h_{n_K}\}} \Bigg[ \prod_j 
 \frac{(\mp 1)^{N_j + H_j}(2J_j+1)^{H_j}}{\prod_{n_j=1}^{N_j} n_j^{4h_{n_j}} h_{n_j}!} 
 \Bigg] \nonumber \\
 && \times \frac{V^H T^{3H-4}}{(2\pi)^{3H}}\Phi (M,\mu_1,\ldots,\mu_H)  
\end{eqnarray}
where $\mu_i$ are meant to be the masses of the lumps defined by the partitions
$\hpartj$ and $H = \sum_j H_j$. Since $H \le N$, the first $H-1$ random numbers, out 
of $N-1$ extracted, can be used to estimate the $\Phi$ integral, according to 
Eq. (\ref{mcphi}), for each term in Eq. (\ref{clusexp3}). 
%--------------------------------------------------------------------------------
\begin{figure}
\resizebox{0.5\textwidth}{!}{
  \includegraphics{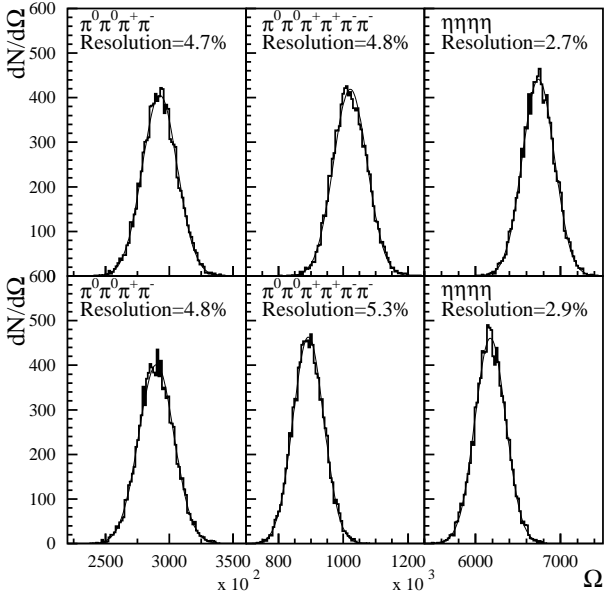}
}
\caption{Distribution of $\Omj$ values computed with a numerical Monte-Carlo 
integration based on 1000 random samples, along with gaussian fits, for a cluster 
of 4 GeV mass and 0.4 GeV/fm$^3$ energy density. Three channels are shown: (a) 
$\pi^0,\pi^0,\pi^+,\pi^-$; (b) $\pi^0,\pi^0,\pi^+,\pi^+,\pi^-,\pi^-$; (c) $\eta,\eta,\eta,
\eta$. Top plots refer to the full quantum statistics calculation while those below to
the classical statistics approximation.}
\label{omega}
\end{figure}
%---------------------------------------------------------------------------------

For $N_S=1000$ random samples, the accuracy of the Monte-Carlo integration is 
satisfactory for all channels and turns out to be of the order of some percent.
This may not be sufficient if one is interested in accurate calculations of single 
channel rates themselves, but is very good for the estimate of other, more inclusive, 
averages such as mean multiplicities, multiplicity distributions etc., in which 
the errors on different channels add incoherently (see Eq. (\ref{obs})) and the 
relative statistical error is therefore greatly reduced with respect to that on 
the single term $\Omega_{\Nj}$. In fig. \ref{omega} we show the gaussian distributions 
obtained by repeating 10000 times the calculation of $\Omj$ for three different 
channels in a cluster with 4 GeV mass and 0.4 GeV/fm$^3$ energy density. 
\footnote{Henceforth energy density must be understood as that in the cluster's 
rest frame, that is the ratio between mass and proper volume $M/V$. The standard
value of 0.4 GeV/fm$^3$ has been chosen and used throughout this paper as it 
corresponds, in the thermodynamical limit, to the energy density of a hadron gas 
at a temperature of about 160 MeV, which has been determined in previous analyses 
of particle production in high energy collisions \cite{becabiele}.}
The plots on the top row refer to the $\Omj$ value calculated in full quantum 
statistics, according to Eq. (\ref{clusexp}), and those on the bottom row in 
classical statistics, i.e. retaining only the leading term~(\ref{boltz}). It can 
be seen that the central values in full quantum statistics are higher than their 
classical approximations, as expected for channels involving identical bosons.
The calculation in full quantum statistics with the exact expression (\ref{clusexp0}) 
for finite volume yields results very close to those obtained with the 
approximated one (\ref{clusexp}).  
The spread in value of the statistical error (see resolutions quoted in 
fig. \ref{omega}) is owing to the different multiplicities and masses of the 
particles in the channels. However, this spread is fairly small and stays well 
below an order of magnitude so that, in fact, a fixed number of Monte-Carlo 
samples $N_S$ is appropriate to calculate most $\Omj$'s with a given accuracy 
independently of particle content.   
%----------------------------------------------------------------------------------
\begin{figure}
\resizebox{0.5\textwidth}{!}{
  \includegraphics{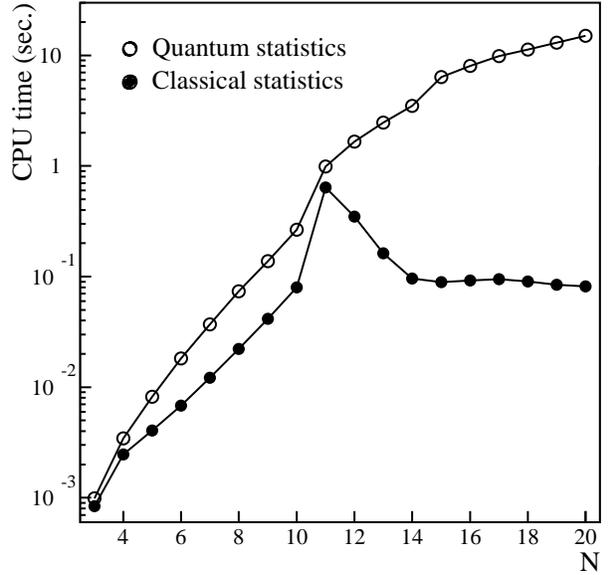}
}
\caption{CPU time needed to calculate the phase space volume of a channel with
$N$ neutral pions in a cluster with mass $N(m_{\pi^0} + 500 \; {\rm MeV})$ and 
energy density 0.4 GeV/fm$^3$.
The full dots refer to classical statistics and the hollow dots to full quantum
statistics; lines are drawn to guide the eye. The CPU time has been normalized 
to a Pentium IV processor with 2 GHz clock rate.}
\label{timechann}
\end{figure}
%----------------------------------------------------------------------------------

The actual CPU time\footnote{All CPU times quoted throughout are referred to a
personal computer with Pentium IV processor working at 2 GHz clock rate, with
rated SPECint=426 and SPECfp=304} needed for the calculation of $\Omj$ with 
1000 random samples is reasonably small. This is shown in fig. \ref{timechann} for 
channels with $N$ neutral pions in a cluster with mass $N(m_{\pi^0} + 500 \; 
{\rm MeV})$ and energy density 0.4 GeV/fm$^3$. In a full quantum statistics 
calculation, including lumps up to 5 particles, the CPU time is much larger 
because of the extra terms in the cluster decomposition. The time needed to 
calculate the integral (\ref{boltz}) increases 
almost exponentially with the number of particles $N$ at low $N$, whereas at 
$N = 11$ it features a drop due to a switch in the numerical method (see Appendix A) 
and flattens out thereafter. In quantum statistics, even at high $N$, there are 
many phase space integrals in the cluster decomposition to be calculated with the 
same method as for the classical terms at low $N$, thus the CPU time steadily
increases, though not as steeply as for the classical term alone.     

%**************************************************************************
\section{Importance sampling of microcanonical ensemble}
%**************************************************************************
  
\begin{figure}
\resizebox{0.5\textwidth}{!}{
  \includegraphics{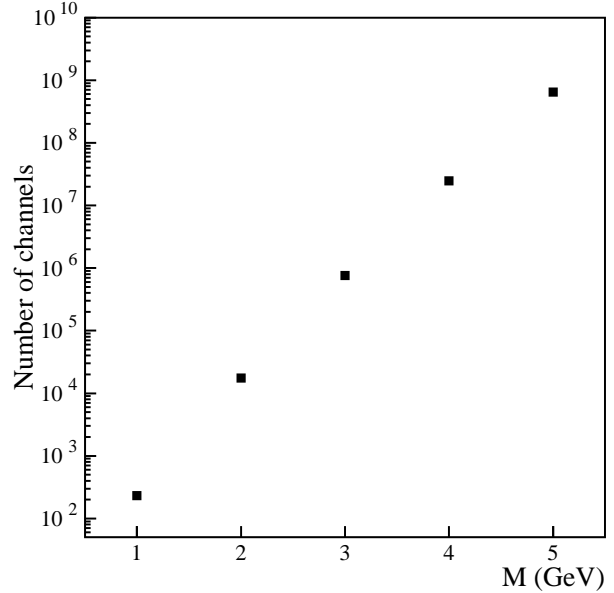}
}  
\caption{Number of allowed channels as a function of cluster mass in a 
hadron gas with 271 species and free charges.}
\label{nchan}
\end{figure}

As has been discussed previously, our goal is to develop a fast and accurate 
numerical method to calculate averages like~(\ref{obs}) in the microcanonical 
ensemble. Being able to effectively calculate $\Omj$ for any channel, a brute 
force option is to do it for all of them. However, this method is not appropriate 
for a system like the hadron gas, because the actual number of channels is huge. 
Indeed, with 271 light-flavoured hadrons and resonances (those included in the 
latest Particle Data Book issue \cite{pdg}), the number of channels allowed by 
energy-momentum conservation is enormous and it increases almost exponentially 
with cluster mass (see fig. \ref{nchan}), involving an unacceptably large 
computing time. For instance, the CPU time needed to compute $\Omj$ for all of 
the 23 millions of channels of a cluster with 4 GeV mass is around 400 hours. 
Charge constraints can indeed reduce significantly the number of allowed channels, 
yet not enough. Therefore, the calculation of the phase space volume of all 
allowed channels is possible only for very light clusters, in practice lighter
than $\approx 2$ GeV.   
%---------------------------------------------------------------------
%---------------------------------------------------------------------
Hence, if a method based on the exhaustive exploration of the channel space 
is not affordable, one has to resort to Monte-Carlo methods, whereby the channel 
space is randomly sampled. 

An estimate of the average (\ref{obs}) can be made by means of the so-called
importance sampling method. The idea of this method is to sample the channel 
space (i.e. the set of integers $N_j$, one for each hadron species) not 
uniformely, rather according to an auxiliary distribution $\Pij$ which must
be suitable to being sampled very efficiently to keep computing time low, 
and, at the same time, as similar as possible to the distribution $\Omj$. The 
latter requirement is dictated by the fact that $\Omj$ is sizeable over a very 
small portion of the whole channel space. Hence, if random configurations were 
generated uniformely, for almost all of them $\Omj$ would have a negligible 
value, thus a huge number of samples would be required to achieve a good accuracy. 
On the other hand, if samples are drawn according to a distribution similar to 
$\Omj$, little time is wasted to explore unimportant regions and the
estimation of the average (\ref{obs}) is more accurate. A crucial requirement 
for $\Pij$ is not to be vanishing or far smaller than $\Omj$ anywhere in its 
domain in order not to exclude some good regions from being sampled, thereby 
biasing the calculated averages in a finite statistics calculation. 

Rewriting Eq. (\ref{obs}) as:
\begin{equation}\label{obs2}
  \langle O \rangle = \frac{\sum_{\Nj} O(\Nj) {\displaystyle \frac{\Omj}{\Pij}} 
  \Pij \delq} {\sum_{\Nj} {\displaystyle \frac{\Omj}{\Pij}} \Pij \delq}
\end{equation}
makes it apparent that a Monte-Carlo estimate of $\langle O \rangle$ 
is:
\begin{equation}\label{estimator}
  \langle O \rangle \doteq \frac{\sum_{k=1}^M O(\Nj^{(k)}) 
  {\displaystyle \frac{\Omj^{(k)}}{\Pij^{(k)}}} }
  {\sum_{k=1}^M {\displaystyle \frac{\Omj^{(k)}}{\Pij^{(k)}}} }
\end{equation}
where $\Nj^{(k)}$ are samples of the channel space extracted according to the 
distribution $\Pi$ and fulfilling the charge constraint $\Qz= \sum_j N_j \qj$.

Provided that $M$ is large enough so that the distributions of both numerator and
denominator in Eq. (\ref{estimator}) are gaussians (hence the conditions of 
validity of the central limit thoerem are met), the statistical error 
$\sigma_\Oav$ on the average $\Oav$ can be estimated as (see Appendix B):
\begin{eqnarray}\label{error}
 \sigma_{\Oav}^2 &=& \frac{1}{M \Omega^2} \Bigg\{ 
 \left[ {\sf E}_\Pi \left( O^2 \frac{\Omj^2}{\Pij^2} \right) - \Oav^2 \Omega^2 
 \right] \nonumber \\
&+& \Oav^2 \left[ {\sf E}_\Pi \left( \frac{\Omj^2}{\Pij^2} \right) - \Omega^2 
 \right] \nonumber \\
&-& 2 \Oav  \left[ {\sf E}_\Pi \left( O \frac{\Omj^2}{\Pij^2} \right)
 - \Oav \Omega^2 \right] \Bigg\}
\end{eqnarray}
where ${\sf E}_\Pi$ stands for the expectation value relevant to the $\Pi$ 
distribution. If $\Omj = \Pij$, the above expression reduces to the familiar
form:
\begin{equation}
 \sigma_\Oav^2 = \frac{1}{M} \left( \langle O^2 \rangle - \Oav^2 \right) 
\end{equation}
The estimator (\ref{estimator}) has a bias unless $\Omj = \Pij$ and this confirms
the necessity to find a distribution as similar as possible to $\Omj$. However, 
the bias, whose general expression is derived in Appendix B, scales with $1/M$ and, 
therefore, for large $M$, can be neglected with respect to the statistical error 
scaling with $1/\sqrt{M}$. 

A possible option for $\Pij$ is the product of $K$ (as many as particle species) 
Poisson distributions:
\begin{equation}\label{poisson}
  \Pij = \prod_{j=1}^K \exp[-\nu_j ] \frac{\nu_j^{N_j}}{N_j!} 
\end{equation}
which will be henceforth referred to as the {\em multi-Poisson distribution}
or MPD. This distribution can indeed be sampled very efficiently and it is 
the actual multi-species multiplicity distribution in the grand-canonical 
ensemble, in the limit of Boltzmann statistics. Although this distribution is 
{\em not} the limit of $\Omj$ in the thermodynamic limit, i.e. for large mass 
and volume of the clusters \cite{goren} (see next section), the similarity between 
MPD and the corresponding microcanonical distribution should be sufficient to 
effectively remove the region of the channel space where $\Omj$ is practically 
vanishing. The mean
values of the MPD are free parameters to be set in order to maximize the similarity
between MPD and $\Omj$. The most sensible choice is to enforce as mean values 
the mean hadronic multiplicities calculated in the grand-canonical ensemble with 
volume and mean energy equal to the volume and mass of the cluster. Indeed, unlike
the higher order moments, the mean multiplicities, or first order moments of the 
multi-species distribution, in the microcanonical ensemble converge to the 
corresponding values in the grand-canonical ensemble in the thermodynamical limit, 
as it will be shown later on. These can be calculated, in the Boltzmann statistics 
by a well known formula:
\begin{equation}\label{canmean}
  \nu_j = \frac{(2J_j+1) V}{2\pi^2} \, m_j^2 T {\rm K}_2 \left(\frac{m_j}{T}\right) 
  \prod_{i} \lambda_i^{q_{ji}}  
\end{equation}
where $V$ is the cluster's volume, $T$ is the temperature and $\lambda_i$ the 
fugacity corresponding to the charge $Q_i$. Temperature and fugacities are determined 
by enforcing the grand-canonical mean energy and charges to be equal to the actual 
mass $M$ and charges $\Qz$ of the cluster:
\begin{eqnarray}\label{saddle}
 &&  M = T^2 \frac{\partial}{\partial T} \sum_j z_j(T) \prod_{i} \lambda_i^{q_{ji}} 
 \nonumber \\
 &&  \Qz = \sum_j \qj z_j(T) \prod_{i}  \lambda_i^{q_{ji}}
\end{eqnarray}
with
\begin{equation}\label{zeta}
  z_j(T) = \frac{(2J_j+1) V}{2\pi^2} \, m_j^2 T {\rm K}_2 \left(\frac{m_j}{T}\right) 
\end{equation}
The (\ref{saddle}) are just the saddle-point equations for the asymptotic expansion 
of the microcanonical partition function, showing that the microcanonical ensemble 
can be approximated by the grand-canonical ensemble for large masses and volumes 
\cite{beca1}. 

It should be stressed that the (\ref{canmean}) is just a particular choice of the
$\nu_j$'s in Eq. (\ref{poisson}), which is by no means compelling. The only purpose 
and merit of the $\nu_j$'s is to make the multiplicity distribution $\Pij$ in 
Eq. (\ref{poisson}) as close as possible to $\Omj$ to speed up the computation. If 
a choice different from Eq. (\ref{canmean}) can do a better job, this could and
should be retained. For the same reason, it makes little sense to use the precise 
canonical expressions of the $\nu_j$'s corrected for quantum statistics, because at
the energy density we are interested in, this is just a correction. Altogether, the 
means in Eq. (\ref{canmean}) turn out to be satisfactory for most practical purposes.   

If cluster charges are unconstrained, only the first of the equations (\ref{saddle}) 
is needed and fugacities can be dropped (i.e. they are taken to be 1) from 
Eq. (\ref{canmean}). On the other hand, if cluster charges are fixed, among all 
configurations drawn from MPD, only those fulfilling charge conservation should be 
retained and considered for microcanonical average calculations in Eq. (\ref{estimator}). 
This preselection of random samples significantly affects the overall efficiency 
of the algorithm, because the acceptance rate of samples extracted from MPD 
becomes small, and it is increasingly smaller for larger clusters. The acceptance 
rate can be improved by resorting to a conditional probability decomposition 
technique, described in the next section.

Summarizing, the procedure to estimate $\langle O \rangle$ in the importance 
sampling method for a given cluster is as follows: 
\begin{enumerate}
\item{} calculate $T$ and $\nu_j$'s according to Eqs. (\ref{saddle}),(\ref{zeta}) 
and (\ref{canmean}),;
\item{} sample the MPD (\ref{poisson}) some large number of times and for each
sample $\Nj$ compute numerically the integral $\Omj$ by the method described in 
Sect.~3 with a suitable number of Monte-Carlo extractions $N_S$;
\item{} calculate the sums in Eq. (\ref{estimator}) and estimate the statistical
error according to Eq.~(\ref{error}).
\end{enumerate}
An improvement in the accuracy of the estimation of $\langle O \rangle$ can be 
obtained by drawing random samples in an extended space instead of calculating 
$\Omj$ for each channel at each step. The idea is to perform the Monte-Carlo 
importance sampling in the variables: 
\begin{equation}
   \{ N_1, \ldots, N_K | r_1, \ldots, r_{N-1} \}  \qquad {\rm with} \;\; N = \sum_j N_j
\end{equation}
at the same time, instead of calculating $\Omj$ separately for each extracted 
sample in $\Nj$. We first note that Eq. (\ref{clusexp3}) can be further rewritten 
as, by using Eq. (\ref{Upsilon}):
\begin{eqnarray}\label{Psi}
 && \Omega_{\Nj} = \!\!\!\!\! \sum_{\{h_{n_1}\},\ldots,\{h_{n_K}\}} 
 \Bigg[ \prod_j \frac{(\mp 1)^{N_j + H_j}(2J_j+1)^{H_j}}
 {\prod_{n_j=1}^{N_j} n_j^{4h_{n_j}} h_{n_j}!} \Bigg] \nonumber \\
 && \times \int_0^1 \d r_1 \ldots \int_0^1 \d r_{H-1} \; \frac{V^H T^{3H-4}}{(2\pi)^{3H}}
 \, \Upsilon (r_1,\ldots,r_{H-1}) \nonumber \\
 && = \int_0^1 \d r_1 \ldots \int_0^1 \d r_{N-1} \!\!\!
 \sum_{\{h_{n_1}\},\ldots,\{h_{n_K}\}} \nonumber \\
 && \Bigg[ \prod_j \frac{(\mp 1)^{N_j + H_j}(2J_j+1)^{H_j}}
 {\prod_{n_j=1}^{N_j} n_j^{4h_{n_j}} h_{n_j}!} \Bigg] 
 \frac{V^H T^{3H-4}}{(2\pi)^{3H}}\, \Upsilon (r_1,\ldots,r_{H-1}) \nonumber \\
 && \equiv \int_0^1 \d r_1 \ldots \int_0^1 \d r_{N-1} \; \Psi ( \Nj | \ri )
\end{eqnarray}
being $\ri \equiv (r_1,\ldots,r_{N-1})$. In practice, each term in the sum has 
been multiplied by $1= \int_0^1 \d r_{H}\ldots \int_0^1 \d r_{N-1}$, 
with $H \le N$ and, doing so, we have been able to take out the integration on the
$r_i$'s. The Monte-Carlo estimate of $\Omj$ as expressed by the last integral 
in Eq. (\ref{Psi}) can then be written as:
\begin{equation}\label{mcestim}
 \Omega_\Nj \doteq \frac{1}{N_S} \sum_{k=1}^{N_S} \Psi (r_1^{(k)},\ldots,r_{N-1}^{(k)})
\end{equation}
where $r_i^{(k)}$ are random numbers uniformely distributed in the interval $[0,1]$.
Likewise, looking at Eqs.~(\ref{obs}) and (\ref{Psi}), the estimator of 
the mean value of the observable $O$ in this extended sampling space can be rewritten 
as:
\begin{equation}\label{gnrlzd}
  \langle O \rangle \doteq \frac{\sum_{k=1}^{M'} O(\Nj^{(k)}) 
  {\displaystyle \frac{\Psi(\Nj^{(k)}|\ri^{(k)}) }{\Pij^{(k)}}} }
  {\sum_{k=1}^{M'} {\displaystyle \frac{\Psi(\Nj^{(k)}|\ri^{(k)})}
  {\Pij^{(k)}}} }
\end{equation}
For a fixed number of calls to the random number generator, this method optimizes
the accuracy because most of them are spent to calculate $\Omj$ for most probable
channels, i.e. those for which an improvement in accuracy is more rewarding,whereas, 
in the previous method, $\Omj$ was calculated with a fixed number of random samples 
$N_S$ regardless of its size. More specifically, if in the previous approach 
$N_S \times M$ computation of the function $\Psi$ in Eq. (\ref{Psi}) were performed, 
about the same CPU time is needed to calculate $M' = N_S \times M$ random samples 
in Eq. (\ref{gnrlzd}), but with a sizeable reduction of the statistical error on 
$\langle O \rangle$. 

In our calculation, resonances whose width exceeds 1 MeV are handled as particles 
with a distributed mass. For each random sample, the function $\Psi$ in Eq. (\ref{Psi}) 
is calculated by randomly drawing masses from a relativistic Breit-Wigner distribution 
for each resonance:
\begin{equation}
  BW(m^2) \d m^2 \propto \frac{1}{(m^2-m_0^2)^2 + \Gamma^2 m_0^2} \, \d m^2
\end{equation}
The mass range is symmetric around the central value $m_0$ with half-width 
$\min\{2\Gamma,m_{cut}\}$ where $m_{cut}$ is the minimal mass allowed by the known 
decay channels of the resonance. If the sum of the extracted masses of particles 
and resonances exceeds the cluster mass, the function $\Psi$ is set to zero.

%*******************************************************************************
\section{Comparison between microcanonical and canonical ensemble}
%*******************************************************************************

The importance sampling method with MPD provides a suitable technique to calculate 
averages in the microcanonical ensemble. In this section we will mainly show 
numerical results on the difference between averages in the microcanonical and 
canonical ensemble.

We have calculated average multiplicities in a hadron gas including 271 light-flavoured 
hadron species up to a mass of about 1.8 GeV quoted in the 2002 Particle Data 
Book issue \cite{pdg}, for completely neutral clusters ($\Qz=0$) and pp-like 
clusters, i.e. with net electric charge $Q=2$, net baryon number $B=2$ and 
vanishing net strangeness. The energy density in the rest frame of the cluster 
$M/V$ has been set to 0.4 GeV/fm$^3$, corresponding, in the thermodynamical limit 
at vanishing chemical potentials, to the temperature value of about 160 MeV found 
in analyses of particle multiplicities in high energy collisions \cite{becabiele}. No extra 
strangeness suppression factor has been used here, as we are just interested 
in a comparison between calculated multiplicities in two different ensembles. 

Whilst the energy density was kept fixed, the mass has been varied from 2 to 14
GeV for neutral cluster and from 4 to 14 GeV for pp-like clusters in steps of 2
MeV. For each mass, 10$^7$ random samples fulfilling charge conservation (i.e. 
passing charge preselection) have been drawn from the MPD. 

In order to improve the performance of the algorithm and decrease the rejection 
rate at the preselection stage, we have implemented a multi-step extraction 
procedure taking advantage of well known properties of the Poisson distribution.
Instead of extracting all particle multiplicities independently from Poisson
distributions, we extracted first the number of baryons $N_B$ and antibaryons 
$N_{\bar B}$ from two Poisson distributions, with means equal to the sums of all 
baryon and antibaryon means respectively, and accept the event only if $N_B - 
N_{\bar B} = B$. Indeed, denoting by $\pi_j$ a single-species Poisson distribution, 
the MPD constrained with charge conservation can be written as:
\begin{eqnarray}
   && \prod_{j=1}^K \pi_j(N_j) \; \delta_{\sum_j N_j \qj,\Qz} = \nonumber \\
 = &&\prod_{{\rm bar}} \pi_j(N_j) \!\!\!
 \prod_{{\rm antibar}} \!\!\! \pi_j(N_j) \!\!\! \prod_{{\rm mesons}}  
 \!\!\! \pi_j(N_j) \; \delta_{\sum_j N_j \qj,\Qz} \nonumber \\
 = && \pi_B(N_B) \pi_{\bar B}(N_{\bar B}) P(N_1,N_2,\ldots | N_B) 
 P(N_{\bar 1},N_{\bar 2},\ldots | N_{\bar B})\nonumber \\ 
 \times && \prod_{{\rm mesons}} \!\!\! \pi(N_j) \; \delta_{N_{B}-N_{\bar B},B} 
 \, \delta_{\sum_j N_j S_j,S} \, \delta_{\sum_j N_j Q_j,Q}
\end{eqnarray}
Here the probability distribution $\prod_{{\rm bar}} \pi(N_j)$ of having given
baryon multiplicities has been decomposed into the product of a Poisson distribution
$\pi_B$ for the overall baryon multiplicity $N_B$ and the conditional probability 
$P(N_1,N_2,\ldots | N_B)$ of having the same set of baryon multiplicities provided 
that their sum is $N_B$; similarly for antibaryons. The $P$ distribution is actually
a multinomial distribution:
\begin{equation}
  P(N_1,N_2,\ldots | N_B) \propto N_B! \prod_j \frac{\nu_j^{N_j}}{N_j!}
\end{equation}
where $\nu_j$ are given by Eq. (\ref{canmean}).
Once all baryon and antibaryon multiplicities have been extracted according to
the multinomial distributions, which can be sampled as quickly as Poisson's, 
the same procedure is carried out for strange mesons. 
Since strange and antistrange mesons are independent of the previous baryon 
extraction, the number of strange mesons $N_S$ and antistrange mesons $N_{\bar S}$ 
are extracted according to a Poisson distribution subject to the requirement that 
$N_S-N_{\bar S} = S - S_B$, where $S_B$ is the net overall strangeness carried by 
the previously extracted baryons. If $N_S$ and $N_{\bar S}$ fulfill the above 
condition, strange and antistrange meson multiplicities are extracted from a 
multinomial distribution, otherwise the event is rejected and the whole 
extraction procedure gets back to the very beginning, i.e. by randomly sampling 
baryon and antibaryon numbers. Likewise, the number of charged non-strange mesons 
and their antiparticles is extracted and, provided that charge conservation is 
fulfilled, their individual multiplicities are determined. Finally, the 
multiplicities of completely neutral mesons, which do not affect the total charges, 
are extracted. The advantage of this method over that based on straight multi-Poisson
sampling can be more easily understood in terms of rejection rate. In fact, the 
number of rejected channels $\Nj$ out of those sampled because of the charge 
constraints, should be the same in both methods as they are drawn from the same 
distribution, i.e. the MPD. Nevertheless, in the latter multi-step method, a 
single rejection of a channel on average does not require $K$ (as many as hadron 
species) extraction from a Poisson distribution: it may well occur at the first 
stage of baryon number conservation, with only one extraction from a Poisson 
distribution or later, with a number of sampled Poisson or multinomial 
distribution which is still less than $K$. The actual gain in CPU time may be 
dramatic, especially for high mass clusters, where the number of allowed channels 
is huge. For instance, for a pp-like cluster 
of 14 GeV, we have estimated a ratio of 0.025 between the average CPU time needed 
to extract a channel with proper charges in the multi-step improved method and in 
the original method. 

\begin{figure}
\resizebox{0.5\textwidth}{!}{%
  \includegraphics{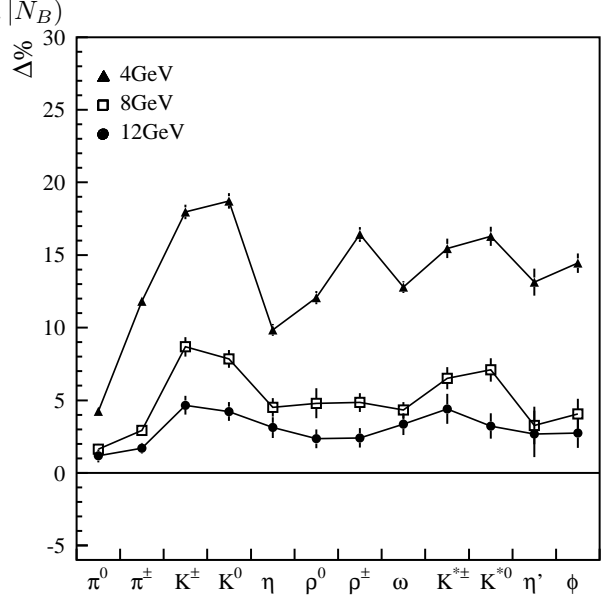}
}
\caption{Relative difference between microcanonical and canonical average 
primary multiplicities of mesons $(\langle N_j \rangle_{\rm micro} - 
\langle N_j \rangle_{\rm can})/\langle N_j \rangle_{\rm can}$ for completely
neutral clusters with mass 4, 8, 12 GeV . The error bars indicate the 
statistical error of the importance sampling Monte-Carlo computation. 
Connecting lines are drawn to guide to eye.}
\label{eemes}
\end{figure}

\begin{figure}
\resizebox{0.5\textwidth}{!}{%
  \includegraphics{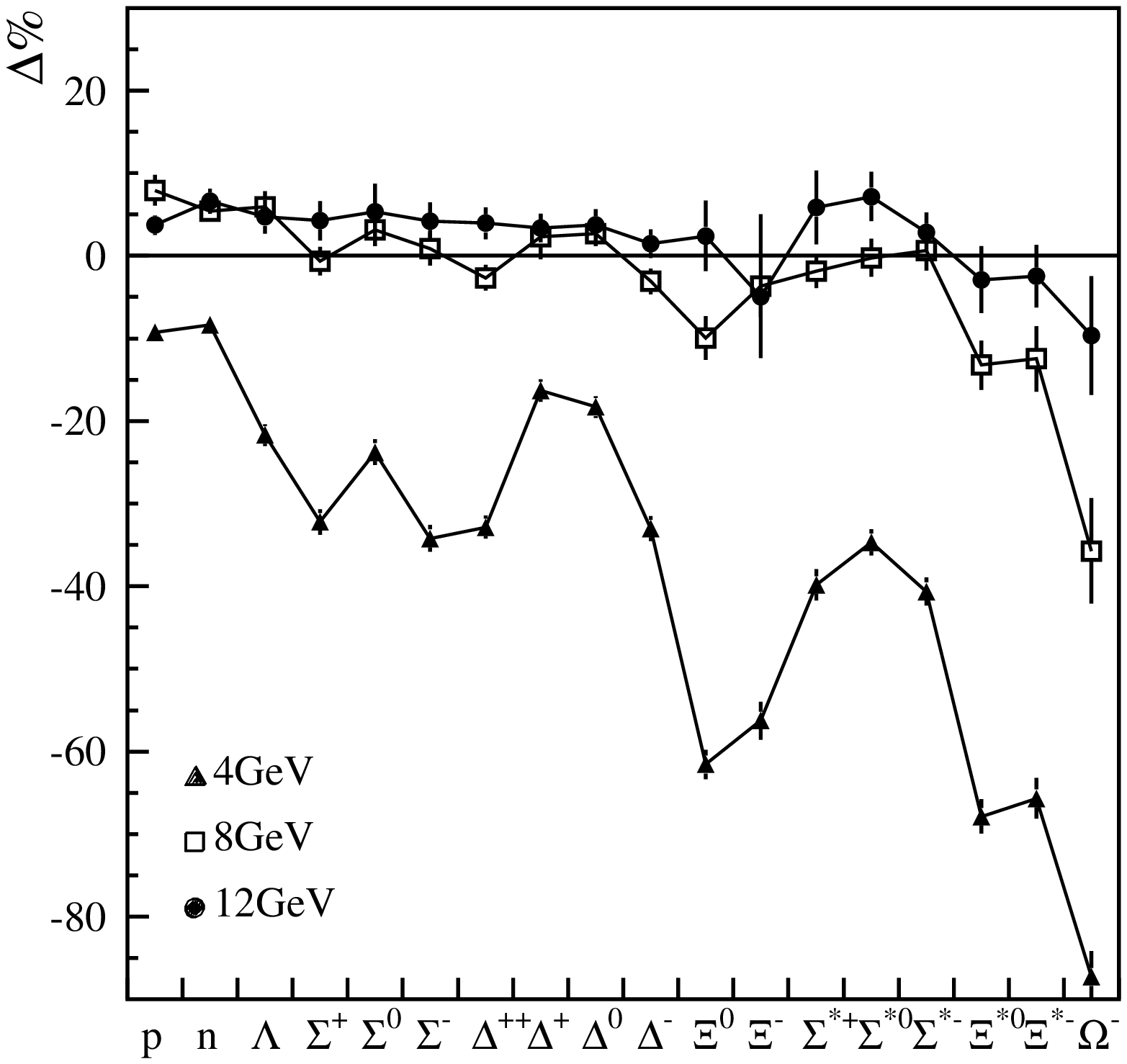}
}
\caption{Relative difference between microcanonical and canonical 
average primary multiplicities of baryons $(\langle N_j \rangle_{\rm micro} - 
\langle N_j \rangle_{\rm can})/\langle N_j \rangle_{\rm can}$ for completely
neutral clusters with mass 4, 8, 12 GeV. The error bars indicate the 
statistical error of the importance sampling Monte-Carlo computation.
Connecting lines are drawn to guide to eye.}
\label{eebar}
\end{figure}

Altogether, the actual CPU time needed to generate 10$^7$ accepted samples (i.e. 
the statistics relevant to the plots in figs. \ref{eemes},\ref{eebar}) of multi-
hadronic channels and calculate average multiplicities in the importance sampling
method for a neutral 4 GeV cluster at 0.4 GeV/fm$^3$ energy density is about 
4.6 10$^2$ sec, i.e. less than 8 minutes.   

The canonical average multiplicities to be compared with the microcanonical
ones have been calculated by determining first the temperature corresponding to
the saddle point equation \cite{beca1}:
\begin{equation}\label{saddle2}
  M - T^2 \frac{\partial}{\partial T} \log Z(\Qz,T) = 0
\end{equation}
where
\begin{eqnarray} 
Z(\Qz,T) &=& \frac{1}{(2\pi)^3} \int_{-\pivs}^{+\pivs} \d^3 \phi \; 
\e^{\i \Qz \cdot \phivs} \exp\Big[\sum_j \frac{(2J_j+1)V}{(2\pi)^3} \nonumber \\
&\times& \int \d^3\p \; \log (1\pm \e^{- \sqrt{\p^2 + m_j^2}/T -\i \qj 
\cdot \phivs})^{\pm 1}\Big] 
\end{eqnarray}
is the canonical partition function; as usual, the upper sign applies to fermions, 
the lower to bosons. Then, multiplicities have been calculated according to the known 
expression in the canonical ensemble \cite{becaheinz}:
\begin{eqnarray}
\!\!\!\!\!\!\!\!\!\!\!\!&& \langle N_j \rangle = \nonumber \\
\!\!\!\!\!\!\!\!\!\!\!\!&& = \sum_{n=1}^\infty (\mp 1)^{n+1} \frac{(2J_j+1)V}{2\pi^2n}
 m_j^2 T{\rm K}_2 \left( \frac{nm}{T} \right) \frac{Z(\Qz - n \qj)}{Z(\Qz)} 
\end{eqnarray} 
where terms in the series beyond $n=1$ account for quantum statistics effects
and are in fact important only for pions at the usually found temperature values of
160-180 MeV. As for the microcanonical ensemble, resonances whose width exceeds
1 MeV have been considered as free hadrons with a mass distributed according to
a relativistic Breit-Wigner distribution.
%-------------------------------------------------------------------------
%-------------------------------------------------------------------------
%-------------------------------------------------------------------------
%-------------------------------------------------------------------------
%-------------------------------------------------------------------------
\begin{figure}
\resizebox{0.5\textwidth}{!}{%
  \includegraphics{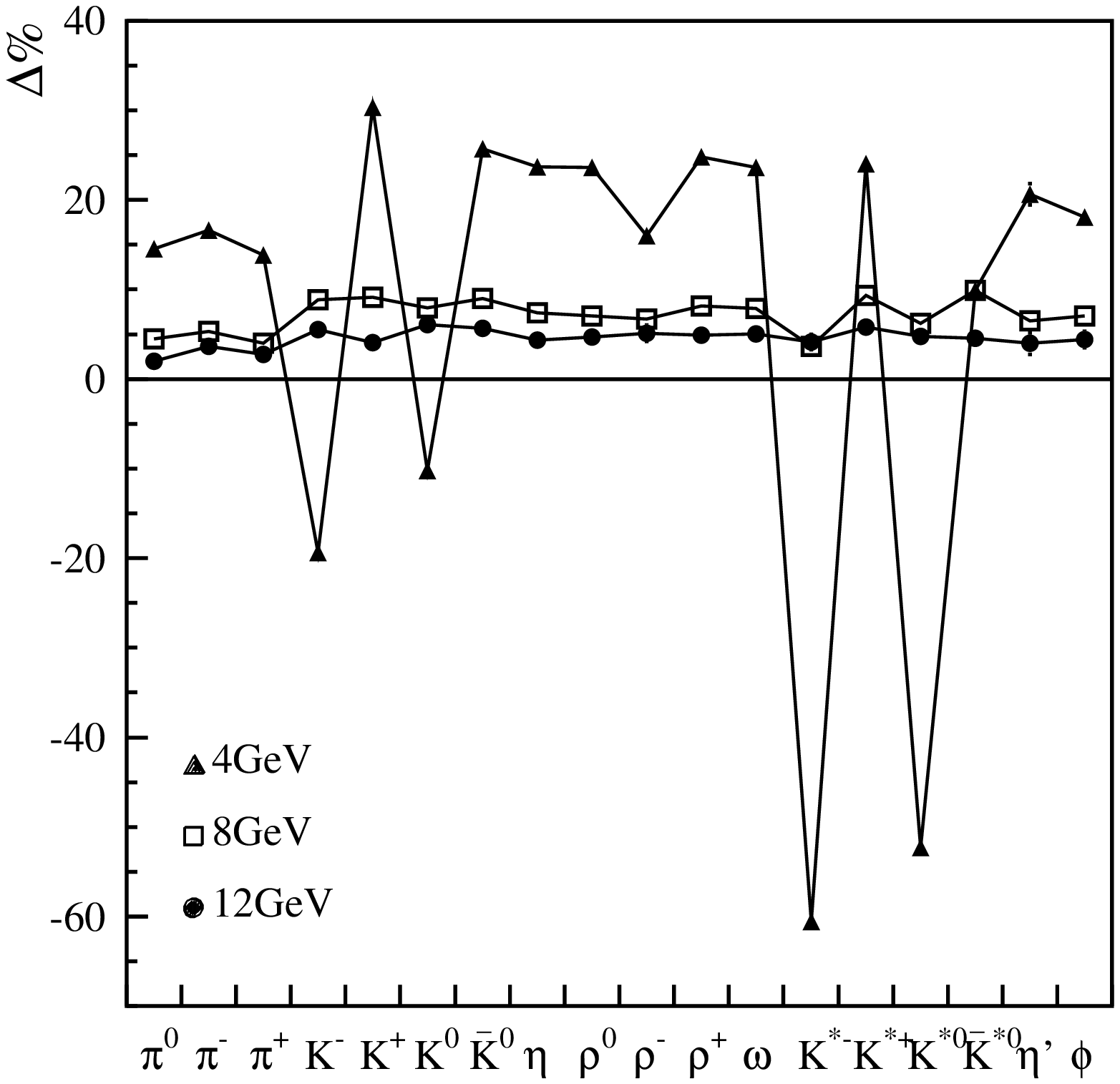}
}
\caption{Relative difference between microcanonical and canonical 
average primary multiplicities of mesons $(\langle N_j \rangle_{\rm micro} - 
\langle N_j \rangle_{\rm can})/\langle N_j \rangle_{\rm can}$ for pp-like 
clusters with mass 4, 8, 12 GeV. The error bars indicate the statistical error 
of the importance sampling Monte-Carlo computation.
Connecting lines are drawn to guide to eye.}
\label{ppmes}
\end{figure}
%-------------------------------------------------------------------------
%-------------------------------------------------------------------------
\begin{figure}
\resizebox{0.5\textwidth}{!}{%
  \includegraphics{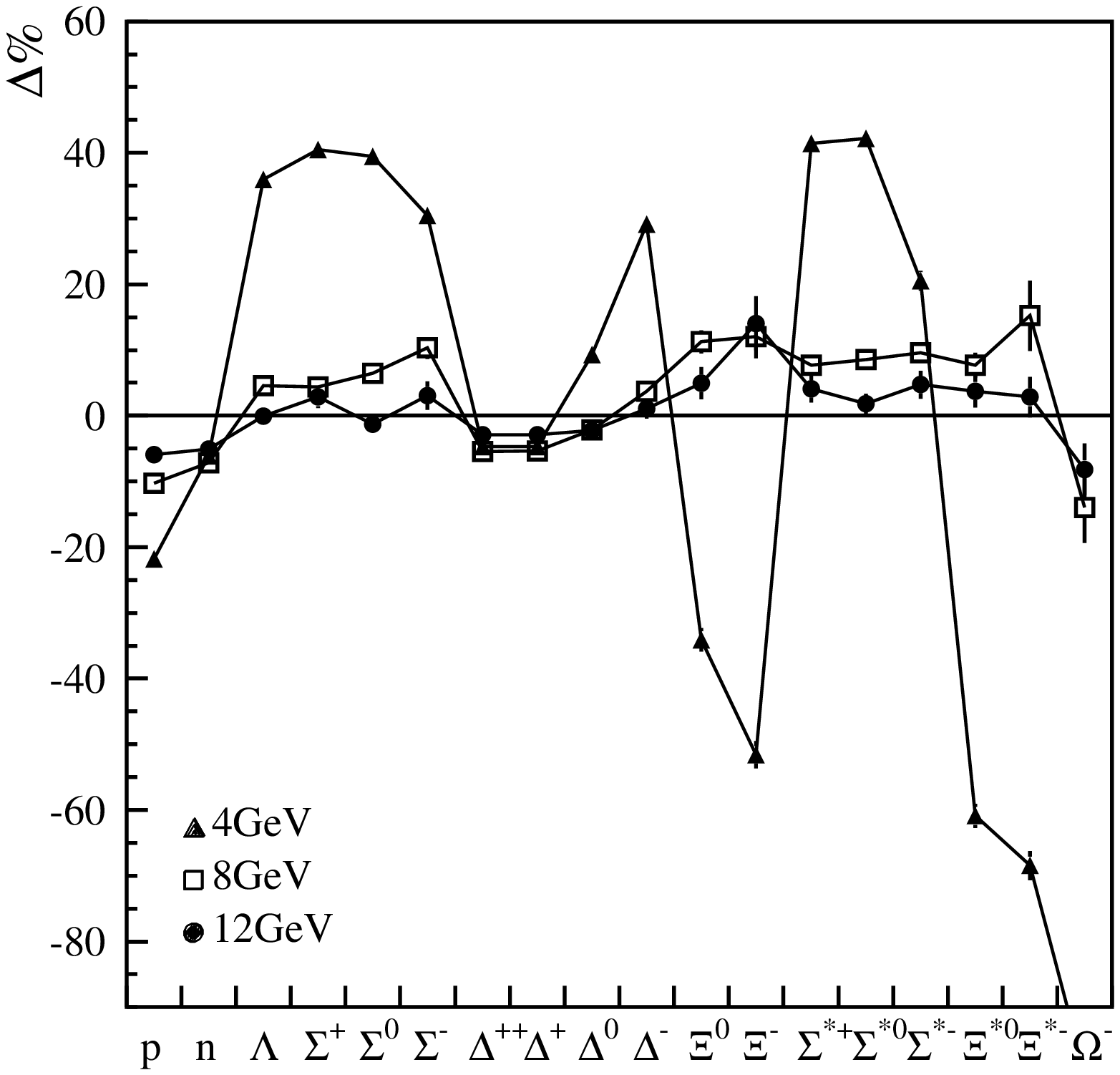}
}
\caption{Relative difference between microcanonical and canonical 
average primary multiplicities of baryons $(\langle N_j \rangle_{\rm micro} - 
\langle N_j \rangle_{\rm can})/\langle N_j \rangle_{\rm can}$ for pp-like clusters 
with mass 4, 8, 12 GeV. The error bars indicate the statistical error of the 
importance sampling Monte-Carlo computation.
Connecting lines are drawn to guide to eye.}
\label{ppbar}
\end{figure}
%-------------------------------------------------------------------------
%-------------------------------------------------------------------------
\begin{figure}
\resizebox{0.5\textwidth}{!}{%
  \includegraphics{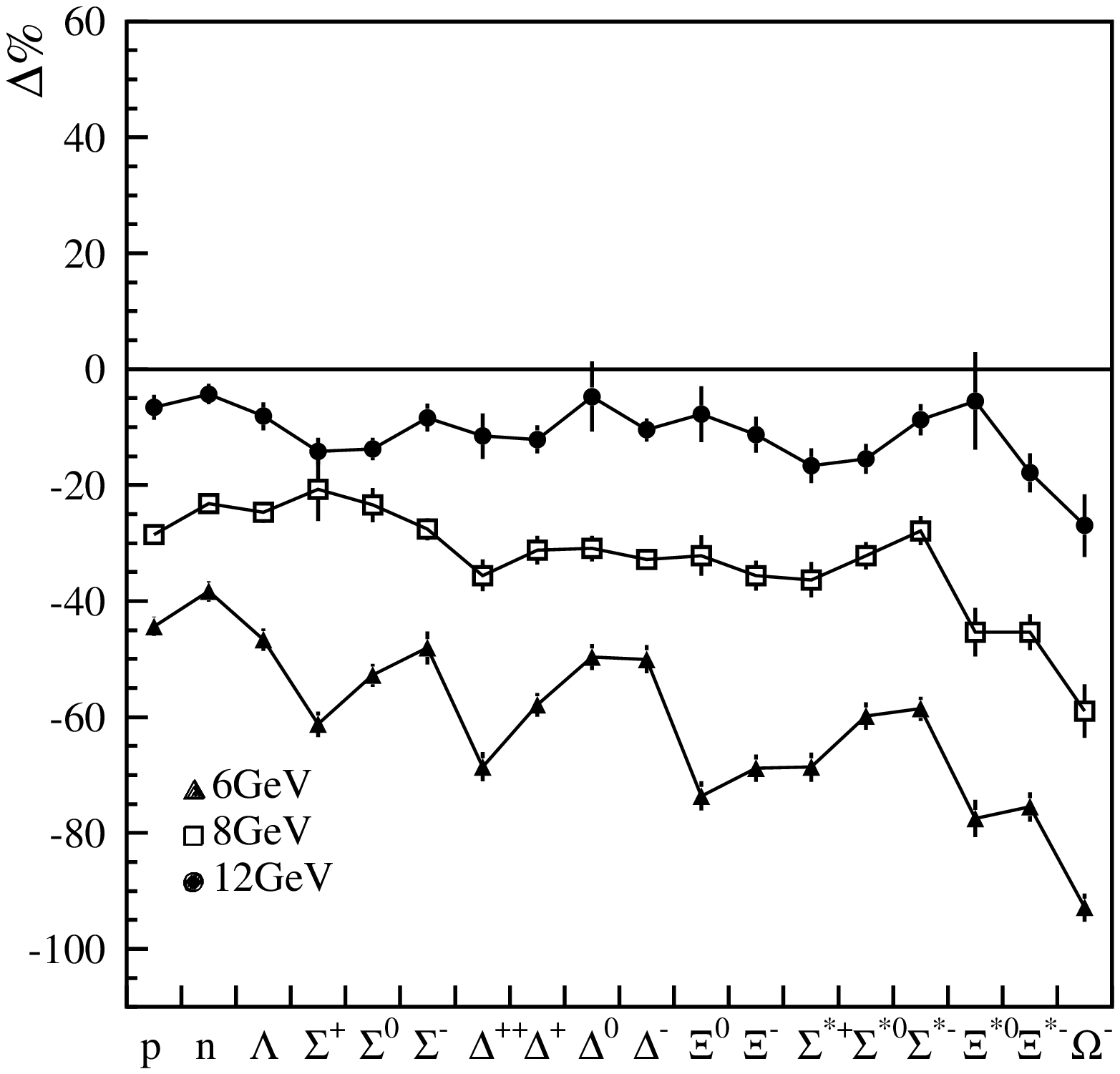}
}
\caption{Relative difference between microcanonical and canonical 
average primary multiplicities of antibaryons $(\langle N_j \rangle_{\rm micro} - 
\langle N_j \rangle_{\rm can})/\langle N_j \rangle_{\rm can}$ for pp-like clusters 
with mass 4, 8, 12 GeV. The error bars indicate the statistical error of the 
importance sampling Monte-Carlo computation.
Connecting lines are drawn to guide to eye.}
\label{ppantibar}
\end{figure}
%-------------------------------------------------------------------------
%-------------------------------------------------------------------------
\begin{figure}
\resizebox{0.5\textwidth}{!}{%
  \includegraphics{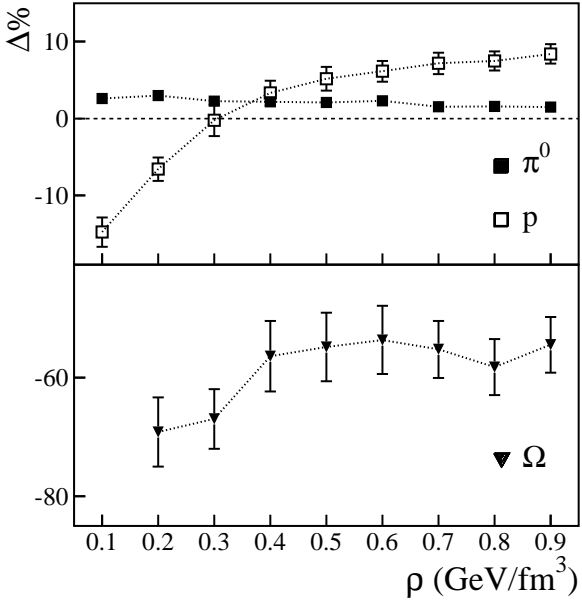}
}
\caption{Relative difference between microcanonical and canonical 
average primary multiplicities of $\pi^0$, p and $\Omega$ 
$(\langle N_j \rangle_{\rm micro} - \langle N_j \rangle_{\rm can})/\langle N_j 
\rangle_{\rm can}$ for a neutral cluster with mass of 6 GeV as a function of
energy density. The error bars indicate the statistical error of the 
importance sampling Monte-Carlo computation. Connecting lines are drawn to guide 
to eye.}
\label{eedens}
\end{figure}
%-------------------------------------------------------------------------

The relative difference between microcanonical and ca\-nonical average for hadrons
belonging to basic SU(3) multiplets and three different cluster masses are shown in 
figs. \ref{eemes},\ref{eebar},\ref{ppmes},\ref{ppbar} and \ref{ppantibar}. At a 
mass of 4 GeV, the deviation is significant and shows a considerable variation as 
a function of the species.
In neutral clusters, mesons and baryons feature a different behaviour: whilst 
microcanonical multiplicities of mesons are higher than the corresponding 
multiplicity in the canonical ensemble, those of baryons are lower. In pp-like
cluster, on the other hand, the general behaviour is not as simple; different
mesons show different signs of the difference and there are even oscillations
of it going from low to high mass clusters. On the other hand, it is evident that 
already at 8 GeV of mass, where the total primary multiplicity of particles is 
around 8 (see table 2), all differences between the microcanonical and canonical 
ensemble do not generally exceed 20\% and further shrink to about 10\% at 12 GeV.
Furthermore, these differences depend only weakly on the energy density, as
shown in fig. \ref{eedens}. We can conclude that, as a rule of thumb, the canonical 
ensemble is a good approximation of the microcanonical one for masses $\gtrsim 8$ 
GeV at energy densities between 0.1 and 0.9 GeV/fm$^3$ with quantum numbers
not greater than that of an elementary colliding system.
Provided that the further assumption of the equivalence between the actual set of
clusters and an equivalent global cluster (EGC) holds \cite{becagp}, this result 
justifies the use of the canonical ensemble in the analysis of particle 
multiplicities in pp, \ee and other high energy collisions. 
Indeed, the mean masses of the EGC corresponding to the actually fitted temperatures 
and volumes, shown in table \ref{masses}, turn out to be sufficiently large 
for most collisions, with the likely exception of K$^+$p at $\sqrt s = 11.5$ GeV
and \ee at $\sqrt s = 14$ GeV where the mean mass is lower than 7 GeV. At 
$\sqrt s \gtrsim 20$ GeV, the mean mass is larger than 8 GeV, implying that the 
canonical ensemble is a good approximation.
 
%-------------------------------------------------------------------------------
\begin{table*}[htb]
\caption{Fitted parameters temperature, volume and extra strangeness suppression parameter
($\gs$ or the mean value of produced strange quark pairs out of the vacuum \ssb) in 
elementary collisions in the canonical analysis of hadron abundances. The canonical fit 
assumes the equivalence, as far as particle multiplicities are concerned, between the 
set of actual clusters and one global cluster \cite{becagp} whose resulting mean mass 
is quoted in the last column. 
Note that \eecc and \eebb events, where heavy flavoured hadrons are emitted, have 
been excluded to estimate $\langle M \rangle$ in \ee collisions.}
\begin{center}
\begin{tabular}{|c|c|c|c|c|c|c|}
\multicolumn{7}{c}{}\\ 
\hline
 Collision &$\sqrt s$ (GeV)&  Reference   &  $T$ (MeV)   &  $V$ (fm$^3$) & $\gs$ or \ssb         &  $\langle M \rangle$ (GeV) \\
\hline
 K$^+$p    &     11.5     & \cite{becagp}& 176.9$\pm$2.6 & 8.12$\pm$0.83 & \ssb = 0.347$\pm$0.020 &   6.51  \\
 K$^+$p    &     21.7     & \cite{becagp}& 175.8$\pm$5.6 & 12.0$\pm$2.4  & \ssb = 0.578$\pm$0.056 &   8.47  \\
 $\pi^+$p  &     21.7     & \cite{becagp}& 170.5$\pm$5.2 & 16.7$\pm$3.1  & \ssb = 0.734$\pm$0.049 &   8.23  \\
 pp        &     17.2     & \cite{bgkms} & 187.2$\pm$6.1 & 6.79$\pm$1.6  & \ssb = 0.381$\pm$0.021 &   7.74  \\
 pp        &     27.4     & \cite{becagp}& 162.4$\pm$5.6 & 25.5$\pm$1.8  & \ssb = 0.653$\pm$0.017 &   9.67  \\
 \ee       &     14       & \cite{becagp}& 167.3$\pm$6.5 & 15.9$\pm$4.1  & $\gs$= 0.795$\pm$0.088 &   6.08  \\   
 \ee       &     22       & \cite{becagp}& 172.5$\pm$6.7 & 15.9$\pm$4.8  & $\gs$= 0.767$\pm$0.094 &   8.12  \\    
 \ee       &     29       & \cite{becagp}& 159.0$\pm$2.6 & 33.1$\pm$4.0  & $\gs$= 0.710$\pm$0.047 &   9.28  \\    
 \ee       &     35       & \cite{becagp}& 158.7$\pm$3.4 & 33.7$\pm$5.2  & $\gs$= 0.746$\pm$0.040 &   9.54  \\    
 \ee       &     43       & \cite{becagp}& 162.5$\pm$8.1 & 29.0$\pm$9.2  & $\gs$= 0.768$\pm$0.065 &   9.99  \\    
 \ee       &     91.25 & \cite{becabiele}& 159.4$\pm$0.8 & 52.4$\pm$2.2  & $\gs$= 0.664$\pm$0.014 &   16.0  \\   
 \ppb      &     200   & \cite{becabiele}& 175$\pm$11    & 35$\pm$14     & $\gs$= 0.491$\pm$0.044 &   21.6  \\   
 \ppb      &     546   & \cite{becabiele}& 167$\pm$11    & 65$\pm$27     & $\gs$= 0.526$\pm$0.044 &   28.7  \\   
 \ppb      &     900   & \cite{becabiele}& 167.6$\pm$9.0 & $\simeq 77$   & $\gs$= 0.533$\pm$0.054 &   34.8  \\   
\hline
\end{tabular}
\label{masses}
\end{center}
\end{table*}
%-------------------------------------------------------------------------------

\begin{figure}
\resizebox{0.5\textwidth}{!}{%
  \includegraphics{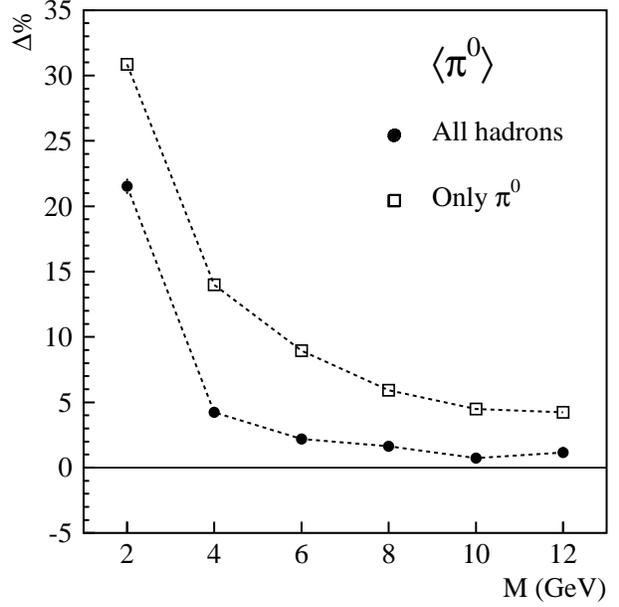}
}
\caption{Relative difference between microcanonical and canonical 
average primary multiplicities of $\pi^0$ $(\langle N_j \rangle_{\rm micro} - 
\langle N_j \rangle_{\rm can})/\langle N_j \rangle_{\rm can}$ a completely
neutral cluster and in a pion gas as a function of cluster mass at an 
energy density of 0.4 GeV/fm$^3$. The error bars indicate the statistical 
error of the importance sampling Monte-Carlo computation.
Connecting lines are drawn to guide to eye.}
\label{pi0gas}
\end{figure}

The quick convergence of microcanonical average multiplicities to canonical 
ones in a hadron gas is favoured by the large number of available degrees of 
freedom. From a mathematical point of view, a large number of degrees of freedom
makes the saddle point expansion converging faster. In physical terms, there 
are more ways to conserve energy and momentum in a system with a larger number
of particle species, so that fulfilling these constraints become less important
earlier than in a system with few degrees of freedom. This is demonstrated in 
fig. \ref{pi0gas} where the relative difference between neutral pion multiplicity
in microcanonical and canonical ensemble is shown for a completely neutral pion
gas and for a full hadron-gas as a function of cluster mass.
%-------------------------------------------------------------------------
%-------------------------------------------------------------------------

\begin{figure}
\resizebox{0.5\textwidth}{!}{%
  \includegraphics{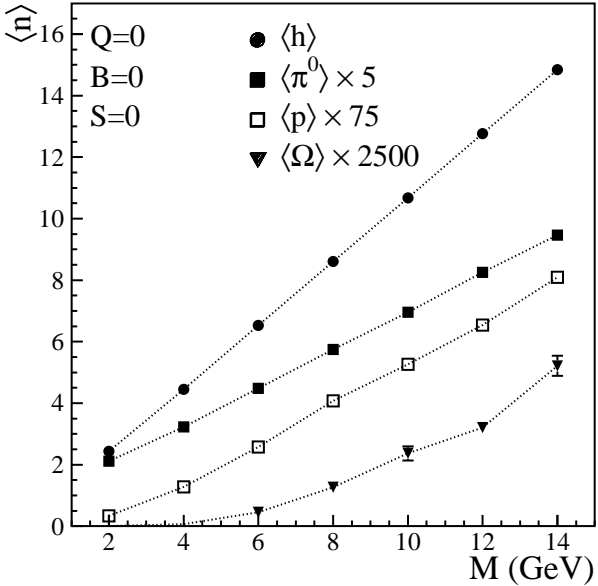}
}
\caption{Microcanonical average primary multiplicities of $\pi^0$, p, $\Omega$
and of all hadron species ($h$) in a completely neutral cluster as a function 
of cluster mass at an energy density of 0.4 GeV/fm$^3$. The error bars indicate 
the statistical error of the importance sampling Monte-Carlo computation.
Lines are drawn to guide to eye.}
\label{multsee}
\end{figure}

The microcanonical average multiplicities of most had\-rons, to a good 
approximation, increase linearly as a function of cluster mass, for fixed 
energy density, starting from $M \simeq 3$ GeV. On the other hand, particles 
with large charge content (such as $\Omega$) show a stronger dependence on mass, 
as shown in fig. \ref{multsee}, over the mass range appropriate for 
microcanonical and canonical calculations. 
%-------------------------------------------------------------------------
%-------------------------------------------------------------------------
  
We have also compared the overall primary multiplicity distributions in the
two ensembles. The multiplicity distribution in the microcanonical ensemble
has been determined by taking $\delta_{N,\sum_j N_j}$ as observable in 
Eqs.~(\ref{obs}),(\ref{obs2}) and performing an importance sampling Monte-Carlo 
calculation.
The multiplicity distribution in the canonical ensemble has been determined
by the same method, that is extracting particle numbers from the MPD and 
weighing each event fulfilling charge conservation with the ratio between
the actual multi-species multiplicity distribution and the MPD. The former
reads (see Appendix C):
\begin{eqnarray}\label{candis}
\!\!\!\!\!\!\! && P({\Nj}) = \frac{1}{Z(\Qz)}\nonumber \\
\!\!\!\!\!\!\! && \times \Bigg[ \prod_j \sum_{\hpartj} 
 (\mp 1)^{N_j + H_j} \prod_{n_j=1}^{N_j} \frac{z_{j(n_j)}^{h_{n_j}}}
 {n_j^{h_{n_j}} h_{n_j}!} \Bigg] \delta_{\Qz,\sum_j N_j \qj}
\end{eqnarray}
where $\hpartj$, as usual, denote partitions and:
\begin{equation}\label{zetap}
 z_{j(n_j)} = \frac{(2J_j+1)V}{(2\pi)^3} \int \d^3\p \; 
 \e^{-n_j \sqrt{\p^2+m_j^2}/T}
\end{equation}
For $T\approx 160$ MeV, only the leading poissonian terms in the distribution
(\ref{candis}) corresponding to $\hpartj=(N,0,\ldots)$ can be retained for 
all particles except pions.

%-------------------------------------------------------------------------
\begin{figure}
\resizebox{0.5\textwidth}{!}{%
  \includegraphics{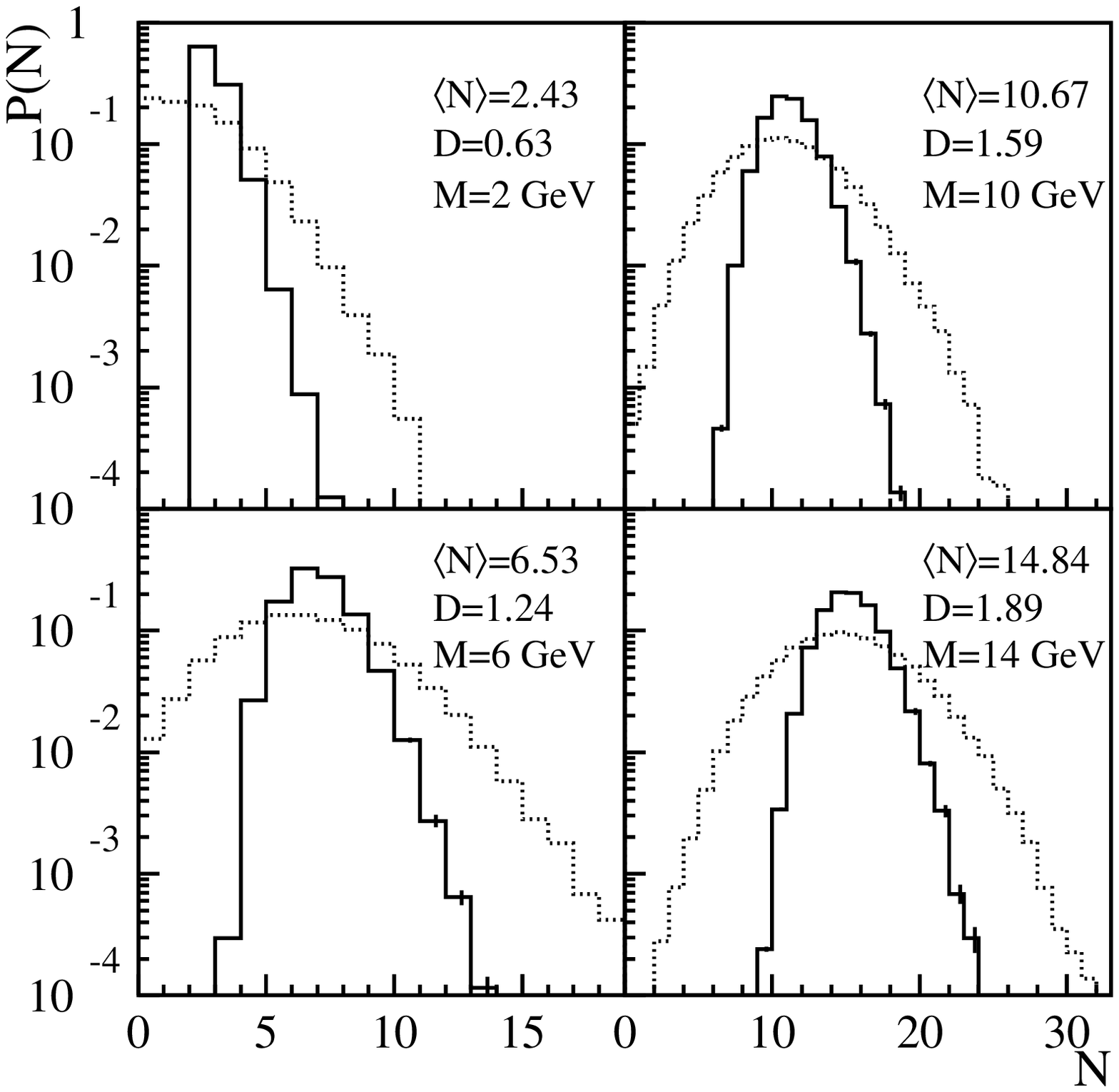}
}
\caption{Comparison between microcanonical (solid) and canonical (dashed) 
overall primary multiplicities distribution in neutral clusters of four 
different masses at an energy density of 0.4 GeV/fm$^3$. }
\label{eedist}
\end{figure}
%-------------------------------------------------------------------------
%-------------------------------------------------------------------------
\begin{figure}
\resizebox{0.5\textwidth}{!}{%
  \includegraphics{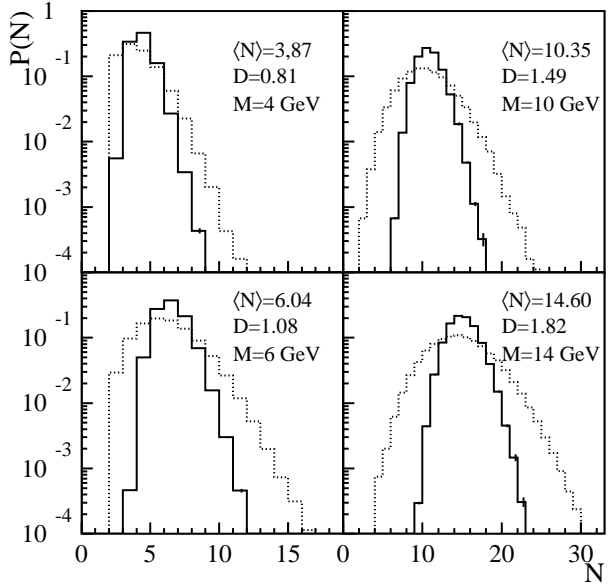}
}
\caption{Comparison between microcanonical (solid) and canonical (dashed) 
overall primary multiplicities distribution in a pp-like cluster of four
different masses at an energy density of 0.4 GeV/fm$^3$. }
\label{ppdist}
\end{figure}
%-------------------------------------------------------------------------

\begin{figure}
\resizebox{0.5\textwidth}{!}{%
  \includegraphics{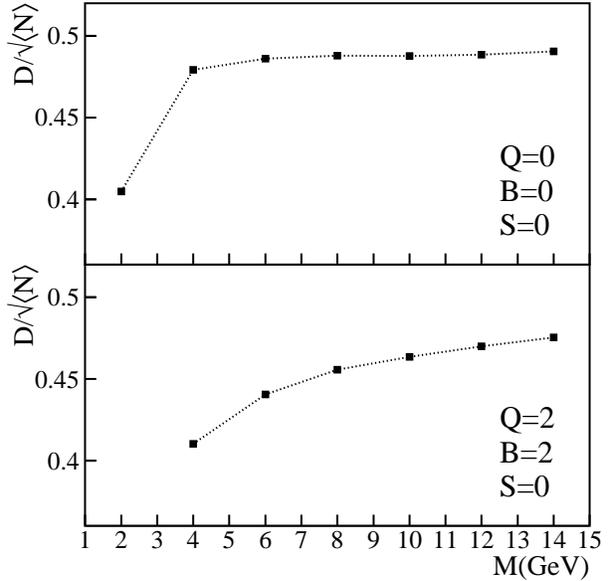}
}
\caption{Ratio between the dispersion of the microcanonical multiplicity 
distribution and the square root of its mean in neutral (top) and pp-like
(bottom) clusters as a function of mass at an energy density of 0.4 GeV/fm$^3$. 
Lines are drawn to guide the eye.}
\label{rappdisp}
\end{figure}

The comparison between the multiplicity distributions are shown in figs. \ref{eedist}
and \ref{ppdist} for neutral and pp-like clusters respectively. It can be
seen that the mean values tend to get closer as mass increases, whilst the
dispersion is lower in the microcanonical than in the canonical ensemble,
where the distribution is almost poissonian. This remarkable effect on particle 
number fluctuation is owing to the overall energy-momentum constraint causing 
a global correlation in particle production. The ratio of the dispersion to
the square root of the mean (i.e. the dispersion of a Poisson distribution)
tends to a factor 1/2 in the thermodynamical limit (see fig. \ref{rappdisp}), 
thus showing that microcanonical ensemble is not equivalent to the grand-canonical 
ensemble with respect to particle number fluctuations. We are not currently aware 
of a simple reason of this fact.  
%-------------------------------------------------------------------------
%-------------------------------------------------------------------------

Because of the persistence of the shape difference between multiplicity 
distributions in the two ensembles, the number of channels sampled with the MPD 
whose weight is much larger than their corresponding microcanonical one increases 
with total mean multiplicity, hence with cluster's mass (at fixed energy density).
This is the reason of the slightly increasing statistical error on average 
multiplicities and multiplicity distributions for increasing cluster mass seen in 
figs. \ref{eemes},\ref{eebar},\ref{ppmes},\ref{ppbar},\ref{ppantibar},\ref{eedist},\ref{ppdist}. 
One could remedy this and speed up microcanonical calculations for very large 
clusters by changing the sampling distribution. For instance, instead of using a MPD 
to sample individual species numbers, the total number of particles can be first 
drawn from a gaussian distribution with variance 1/4 of the sum of all hadron 
primary multiplicities as estimated in the canonical ensemble; then, the number 
of baryons $N_B$, strange mesons $N_S$, charged non-strange mesons $N_Q$, neutral 
mesons $N_0$ and their respective number of antiparticles could be sampled from a 
multinomial distribution:
\begin{equation}
  P = N!\prod_{i=1}^7 \frac{\xi_i^{N_i}}{N_i!}
\end{equation}
where $i=B,{\bar B},S,{\bar S},Q,{\bar Q},0$ and $\xi_i$ are the sums of all
$\nu_j$ functions in Eq. (\ref{canmean}) of particles belonging to the same 
class $i$. Finally, once the multiplicities of each class $N_i$ fulfilling conservation
laws have been extracted, individual particle multiplicities could be determined
for each class in turn by using again multinomial distributions. 
 
%-------------------------------------------------------------------------
\begin{table}[htb]
\caption{Mean multiplicity $\langle N \rangle$ and dispersion $D$ of the 
microcanonical and canonical multiplicity distribution for neutral and pp-like 
clusters as a function of the mass. Also quoted the corresponding
temperature in the canonical ensemble obtained from Eq. (\ref{saddle2}).}
\begin{center}
\begin{tabular}{|l|l|l|l|l|l|}
\hline\noalign{\smallskip}
\multicolumn{6}{c}{Neutral cluster}\\
\noalign{\smallskip}\hline
 &  \multicolumn{2}{|l|}{Microcanonical} & \multicolumn{3}{|l|}{Canonical} \\
\hline 
$M$ (GeV) & $\langle N \rangle$& $D$ & $\langle N \rangle$ & $D$ & $T$ (MeV)\\
\hline
 2	&   2.43    &  0.63  & 1.96  &     1.74  &  175.3 \\
 4	&   4.45    &  1.01  & 4.08  &     2.45  &  169.3 \\
 6	&   6.53    &  1.24  & 6.17  &     2.93  &  166.3 \\
 8	&   8.61    &  1.43  & 8.27  &     3.29  &  164.6 \\
 10	&   10.67   &  1.59  & 10.38 &     3.58  &  163.5 \\
 12	&   12.76   &  1.75  & 12.46 &     3.90  &  162.8 \\
 14	&   14.84   &  1.89  & 14.56 &     4.15  &  162.2 \\
\hline\noalign{\smallskip}
\multicolumn{6}{c}{pp-like cluster}\\
\noalign{\smallskip}\hline
 &  \multicolumn{2}{|l|}{Microcanonical} & \multicolumn{3}{|l|}{Canonical} \\
\hline 
$M$ (GeV) & $\langle N \rangle$& $D$ & $\langle N \rangle$ & $D$ & $T$ (MeV)\\
\hline
 4	&   3.87  &  0.81  & 3.63   &	  1.33 &  142.1  \\
 6	&   6.04  &  1.08  & 5.76   &	  2.05 &  152.5  \\
 8	&   8.20  &  1.30  & 7.90   &	  2.59 &  156.1  \\
 10	&   10.35 &  1.49  & 10.04  &	  3.02 &  157.8  \\
 12	&   12.48 &  1.66  & 12.17  &	  3.40 &  158.7  \\
 14	&   14.60 &  1.82  & 14.27  &	  3.73 &  159.2  \\
\hline 
\end{tabular}
\end{center}
\end{table}
%-----------------------------------------------------------------------------------------

%***********************************************************************
\section{The Metropolis algorithm}
%***********************************************************************

The importance sampling method allows to calculate averages like~(\ref{obs})
in single Monte-Carlo runs quite straightforwardly and can be used as an event 
generator if events are reweighted, as we have seen, by a factor $\Omj/\Pij$. 
If one needs to sample $\Omj$ directly without reweighting the events, different 
methods should be considered. 
A first possibility is a rejection method, but it can be soon realized that it 
is unfit for the problem we are dealing with. In fact, for this method to be 
effective, one needs a covering function $F(\Nj)$ (i.e. a function of the channel 
such that $F(\Nj) > \Omj \; \forall \Nj$ which can be sampled very 
efficiently and as close as possible to the $\Omj$. Such a function is hard to 
find because $\Omj$ is not smooth in its domain; strong variations of 
the phase space volume may occur by changing just one particle. We have seen that 
the MPD is likely to be similar to our target distribution, but in order to be 
a covering function, it should be rescaled by a factor $c$ such that $c \Pij > \Omj 
\; \forall \Nj$. However, to estimate $c$, one ought to evaluate $\Omj$ for all 
channels and this is just what cannot be afforded. A general method to sample 
complex multi-dimensional distributions is the Metropolis algorithm \cite{metro}, 
which has been applied to the specific problem of numerical calculations of the 
multihadronic microcanonical ensemble by Werner and Aichelin \cite{weai}. 
In this work, we will show that this method can be further improved in speed and 
accuracy and, at the same time, that great care is needed in handling it, 
especially when assessing the equilibrium conditions.  

The Metropolis algorithm prescribes the implementation of a random walk in the 
channel space on the basis of acceptance or rejection of proposed transitions
from the current position. After some number of steps, the probability of visiting 
a given point is proportional to the target distribution; otherwise stated, the 
points in the random walk are actual samples of the the target distribution (in 
our case $\Omj$) and they can be stored as generated events. We will now discuss 
more in detail how this comes about.
In a general random walk, the probability $P_m(t)$ of visiting a state $m$ (i.e. 
a channel or multi-hadronic configuration) at the $t^{\rm th}$ step evolves according 
to the master equation:
\begin{equation}\label{master}
  P_m(t+1) - P_m(t) = \sum_{n} P_n(t) w(n\rightarrow m)- P_m(t) w(m\rightarrow n)
\end{equation}
where $w(n\rightarrow m)$ is the probability of transition from the state $n$ to
the state $m$. At the equilibrium:
\begin{equation}\label{equilib}
P_m(t+1)= P_m(t) \qquad \forall m 
\end{equation} 
This occurs if (a sufficient but not necessary condition):
\begin{equation}
  P_n (t) w(n\rightarrow m) = P_m (t) w(m\rightarrow n) \qquad \forall n,m 
\end{equation}
If we want the probabilities $P_m$ to be proportional to a given target distribution
$f(m)$, so that the number of times the state $m$ is visited from some step $t$
onwards is proportional to $f(m)$, we have to enforce the condition:
\begin{equation}\label{balance}
\frac{w(n\rightarrow m)}{w(m\rightarrow n)}=\frac{f(m)}{f(n)} \; .
\end{equation}
Provided that the above equation is fulfilled, the choice of a set of transition 
probabilities $w$ (a so-called {\em updating rule}), is free. The only requirement 
is that every point can be reached from any point in a finite number of steps 
with non-vanishing probability (ergodicity condition), otherwise there would be 
unaccessible regions even though $f \ne 0$ therein. Although any choice of the 
$w$'s is in principle allowed, some are worthier. Indeed, the transition 
probabilities $w$ govern the dynamical behaviour of the random walk and, particularly, 
how fast the system gets to the equilibrium condition~(\ref{equilib}) after a 
transient. This is owing to the fact that, in general, the random walk starts 
from states which are not random samples of the target distribution, so that 
$P_m(0) \ne f(m)$. A good choice of the $w$'s will keep the {\em relaxation time} 
$T_{rel}$, defined as the number of steps needed to get sufficiently close to the 
equilibrium value, to a minimum, thereby making event generation faster.

In general, the transition probability can be decomposed into a proposal probability
$T$ ({\em proposal matrix}), i.e. the probability of considering a given transition,
and the conditional probability $A$ of accepting it once it has been proposed. 
In symbols:
\begin{equation}
\nonumber w(n\rightarrow m)=T(n\rightarrow m) A(n\rightarrow m) \; .
\end{equation}
Starting from a non-equilibrium situation, a physical system reaches equilibrium 
earlier if transition probabilities are larger. Likewise, in the Metropolis 
algorithm, the relaxation time is little if $w(n\rightarrow m)$ are large; in
other words, if $w(n \rightarrow n)$ is small taking into account the normalization 
condition:
\begin{equation}
  \sum_m w(n\rightarrow m) = 1
\end{equation}
For the transition probabilities to states different from the current state
to be large, the acceptance matrix should be as large as possible for $m \ne n$
once the proposal matrix is known. There is indeed an optimal choice for $A$ which 
reads:
\begin{equation}\label{accept}
 A(n\rightarrow m)= \min \Bigg\{1,\frac{f(m)T(m\rightarrow n)}
 {f(n) T(n \rightarrow m)} \Bigg\} \; .
\end{equation}  
This choice maximizes $w(n\rightarrow m)$ if $n \ne m$ \cite{umri} and fulfills 
the condition~(\ref{balance}). The next problem is choose a good proposal matrix. 
The issue is discussed in detail in ref.~\cite{umri} and the conclusion is that 
$T(n \rightarrow m) \simeq f(m)$, i.e. the proposal matrix should be an easy-to-sample 
distribution as close as possible to the target distribution $f(m)$. This can be 
easily understood by taking the limiting case (which, if possible, would make the 
Metropolis algorithm unnecessary):
\begin{equation}\label{limit}
  T(n \rightarrow m) = f(m)
\end{equation}
In this case, the relaxation time would be zero: the starting point, as well as
all other points in the random walk, would be sampled from the target distribution 
itself and any transition would be accepted because $A(n \rightarrow m) = 1$ 
according to Eqs.~(\ref{accept}) and (\ref{limit}). While the condition (\ref{limit})
cannot be obtained in practice (Metropolis algorithm would be unnecessary in 
that case), one can try to get as close as possible to it. From what we have 
seen in Sect.~4 about importance sampling, it can be argued that the MPD in 
Eq. (\ref{poisson}) would be a good proposal matrix and could also be used to 
pick the starting point in the Metropolis random walk. We will see the benefits 
of the use of the MPD more in detail in the next section.

A drawback of the Metropolis algorithm is that, unlike in the importance sampling 
method, different samples (i.e. steps in the random walk) are not independent. 
This can be easily understood reminding that proposed transitions may be rejected, 
so that a point may appear several times in a row in the random walk. Hence, there 
is a finite positive 
statistical correlation between the values of physical observables at different 
steps and this gives rise to an increase of the overall uncertainty in the estimate 
of averages with respect to the case of independent samples. The fact that different 
events are correlated may render the use of Metropolis algorithm not appropriate 
in some applications. However, in the problem of high energy collision event 
simulation, where one has to hadronize many clusters with different masses in 
each event, this is not an issue; in this case, a single Metropolis random 
walk must be run for each cluster, and only one sample, representative of its  
microcanonical ensemble, drawn after equilibrium has been reached.

The Metropolis algorithm can be used to estimate the average of an observable 
like~(\ref{obs}) in the microcanonical ensemble of a single cluster by taking
a sufficiently large number of steps (i.e. $ \gg T_{rel}$) in one random walk and 
calculating:
\begin{equation}\label{metroestim}
 \Oav \doteq \frac{\sum_{k=1}^M O^{(k)}}{M}
\end{equation}
where $M$ is the total number of steps and $O^{(k)}$ the actual value of the 
observable $O$ at $k^{\rm th}$ step. A nice feature of the Metropolis algorithm 
is that, unlike in the importance sampling method, there is no need of overall 
normalization when estimating $\Oav$ (compare Eq. (\ref{metroestim}) with 
Eq. (\ref{estimator})). On the other hand, as already emphasized, the values of 
$O$ at different steps are correlated. The statistical error on $\Oav$ in 
Eq. (\ref{metroestim}) can be estimated as (see Appendix D):
\begin{equation}\label{metroerro}
   \sigma_\Oav = \sqrt{\frac{\langle O^2 \rangle - \Oav^2+2R}{M}} 
\end{equation}
where $R$ is the integral of the {\em autocorrelation function} $A$, defined here as 
the difference between the expectation value of the product of the observable value at 
the steps $k$ and $k+h$ and its average $\langle O \rangle^2$
\begin{equation}
 A(h) \equiv {\sf E}({\sf O}^{(k)} {\sf O}^{(k+h)}) - \langle O \rangle^2 
\end{equation}  
If the starting point is random, the autocorrelation function is independent of $k$ 
and gives information about how correlated are distant steps, namely how long the 
system keeps memory of its past steps. As already mentioned, this correlation 
between different steps arises from the finite probability of rejecting transitions 
and vanishes only if all proposed transitions are accepted, i.e. if Eq. (\ref{limit}) 
is fulfilled. Therefore, the autocorrelation function is always positive and vanishes 
only if different steps are independent. Moreover, the 
statistical error (\ref{metroerro}) is larger than in the case of uncorrelated 
samples, where it reaches its minimum. The number of steps needed to reduce $A(h)$ 
to some small fraction of $\Oav^2$ is defined as {\em autocorrelation time} $T_{auto}$. 
Thus, in order to minimize the statistical error on $\Oav$ in Eq. (\ref{metroerro}),
one should keep the autocorrelation time as little as possible.

The autocorrelation function and time depend only on the updating rule, whilst the 
relaxation time $T_{rel}$ also depends on how the initial state is chosen. In 
principle, the relaxation time might be zero while the autocorrelation time is 
not. When using the MPD both as proposal matrix and to generate the starting 
point, these two quantities are tightly related. 

\begin{figure}
\resizebox{0.5\textwidth}{!}{
  \includegraphics{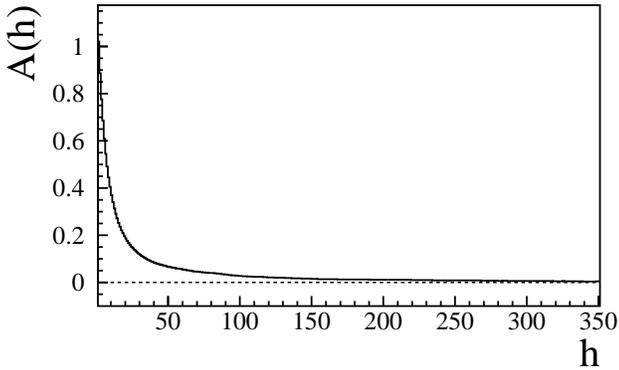}
}  
\caption{Autocorrelation function for the total primary multiplicity for a
neutral cluster of 4 GeV mass at an energy density of 0.4 GeV/fm$^3$.}
\label{autocorr}
\end{figure}

The autocorrelation function can be estimated during the Metropolis random walk 
by the sum:
\begin{equation}
  A(h) \doteq \frac{\sum_{k=1}^{M-T_{auto}} O^{(k)} O^{(k+h)}}{M-T_{auto}} - 
  \frac{1}{M^2}\left(\sum_{k=1}^M  O^{(k)}\right)^2
\end{equation}
and it is shown in fig. \ref{autocorr} for a cluster of 4 GeV with updating
rule based on the MPD.

Now the question arises whether the Metropolis algorithm leads to more or less 
accurate computations of mean values than the importance sampling method, for 
given computing resources and using the same distribution $\Pij$ as proposal matrix 
and sampling distribution respectively. This is studied in more detail in 
Subsect.~7.4.

%---------------------------------------------------------------------
%---------------------------------------------------------------------

%***********************************************************************
\section{Study of the Metropolis algorithm}
%***********************************************************************    
    
We have studied the capability of the Metropolis algorithm as an event generator 
for the statistical model of hadronization and as a computing tool for the hadron 
gas microcanonical ensemble. As has been discussed in the previous section, the 
benchmark for this algorithm is the number of steps needed to reach equilibrium, 
or relaxation time $T_{rel}$, which must be kept to a minimum so as to draw 
samples of the target distribution as quickly as possible. Likewise, the 
autocorrelation time $T_{auto}$ must be small in order to minimize the statistical
error (\ref{metroerro}) on the estimate of mean values in single random walks.
As has been mentioned at the end of previous section, these two quantities are tightly 
related in our MPD-based scheme, so we can confine ourselves to study $T_{rel}$. 

The relaxation time in principle depends on the physical observable (average 
multiplicities, multiplicity distribution etc.) and it is not easy to 
estimate in advance. Hence, in practice, it must be determined {\em a posteriori}
by analyzing the convergence to equilibrium for the observable of interest. For 
instance, if we ought to calculate the overall multiplicity distribution $P_n$, we 
would have to study the height of the $n^{\rm th}$ bin, for all $n$'s, as a function 
of the step. The height of the $n^{\rm th}$ bin at the $k^{\rm th}$ step, that is 
$P_{n(k)}$, can be estimated by repeating the Metropolis random walk many times 
and averaging:
\begin{equation}\label{metromean}
   P_{n(k)} = \frac{1}{L} \sum_{i=1}^L \delta_{n,n_{i(k)}}
\end{equation}
where $n_{i(k)}$ is the actual multiplicity in the $i^{\rm th}$ random walk at 
the step $k$. 

\begin{figure}
\resizebox{0.5\textwidth}{!}{
  \includegraphics{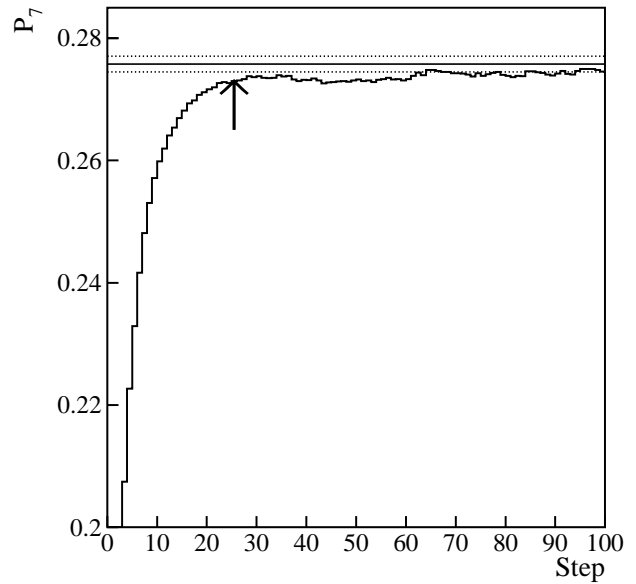}
}  
\caption{Step hystogram showing the convergence to equilibrium of the 
probability $P_{7}$ of a channel with 7 primary hadrons as a function of the 
step in a set of 500000 Metropolis random walks for a neutral cluster of 6 GeV 
with energy density of 0.4 GeV/fm$^3$. The horizontal solid
line indicates the $P_{7}$ value estimated with the importance sampling method
and the dashed band its relevant statistical uncertainty. The arrow points 
to the equilibrium point determined through the WOSSR test (see text).}
\label{step}
\end{figure}

Establishing when a stable value of the examined observable is attained can be
done by studying its average over many Metropolis random walks, like (\ref{metromean}) 
as a function of the step, i.e. forming a hystogram (the {\em step hystogram}, see 
fig. \ref{step}) with Metropolis step numbers as bins and the average of the 
observable (e.g. multiplicity) as bin content. In all considered cases, the step 
hystogram shows the same general behaviour, namely a tendency to an equilibrium 
value after, possibly, a strong initial oscillation and followed, sometimes, by 
very mild damped oscillations. Step hystograms show fluctuations around an apparent 
equilibrium value at some scale, as it can be seen in figs. \ref{step},\ref{step-int}. 
The fluctuation pattern is determined by the interplay of finite statistics and 
possible dynamical oscillations governed by the master equation (\ref{master}). The 
amplitude of statistical fluctuations is a function of the updating rule.
      
%---------------------------------------------------------------------
%---------------------------------------------------------------------
Because of these facts, providing a good quantitative assessment of where 
stability is achieved is not straightforward. Were the statistics of Metropolis
random walks infinite, the criterium for stability would be fully deterministic,
that is based on the relative difference between asymptotic value (the true 
microcanonical average) and actual value from a given step onwards. On the other 
hand, we can only deal with finite statistics, so the estimation of a stability 
point (thence $T_{rel}$) for a given observable can be done statistically, thus 
it will be affected by some uncertainty. 

The statistical method we have chosen in order to assess the convergence to 
equilibrium is based on a non-parametric statistical test, the {\em Wilcoxon 
one sample signed rank test} (WOSSR), that we will shortly describe in the following.
The WOSSR test is a non-parametric test, i.e. it does not require the knowledge of 
the statistical distribution of the sample, and the hypothesis under test (the {\em 
null hypothesis}) is that a set of $M$ independent random samples $\{x_1,\ldots,x_M\}$ 
has a given median $X_m$. The only requirement is that their distribution is 
symmetric. The test procedure consists in sorting the absolute values 
of the differences $d_i=x_i-X_m \;\; \forall i$ and giving them a rank, namely 1 
to the largest, 2 to the second largest and so on. If $X_m$ is a good median, the 
sum of the ranks of the positive differences is expected to be close to the sum 
of the ranks of negative differences. The test statistic is just the sum $W$ 
of the ranks of the positive differences and the test result is good when $W$ 
is close to half the sum of the first $M$ integers, that is $M(M+1)/4$.

We have used this method to study the convergence to equilibrium in the Metropolis 
algorithm taking as random samples a subset of $M$ running bins (from the 
$k^{\rm th}$ to the $(k+M)^{\rm th}$) in the step hystogram of the multiplicity 
distribution $P_{n(k)}$ (see fig. \ref{step}). In fact, if equilibrium is reached 
in the Metropolis random walk (as it is apparent in the rightmost part of the 
hystogram in fig. \ref{step}), $P_{n(k)}$ is expected to evenly fluctuate around 
the asymptotic equilibrium value, whereas a net drift towards this value appears 
when out of equilibrium (i.e. in the leftmost part of fig. \ref{step}).
Therefore, provided that the distribution of fluctuations is symmetric at equilibrium, 
the asymptotic 
value is likely to be a good median for sets of $P_{n(k)}$ with $k=1,\ldots,M$ in
the equilibrium region and not in the drifting region. Accordingly, the WOSSR test 
will yield a good confidence level in the former case and a very small 
one in the latter. The true asymptotic value is not known in practice and must be 
estimated from the step hystogram itself. A good choice is the arithmetic mean of 
a set of hystogram contents in the rightmost region, where stability is apparently 
reached.

The non-parametric WOSSR test carried out on a set of $M$ running bins in the step
hystogram seems suitable for our problem. In fact, we do not know the statistical 
distribution of the random variable ${\sf O}^{(k)}$ (i.e. the examined observable) 
at each step for a given number $L$ of random walks. On the other hand, it is 
reasonable to assume that it is symmetric at equilibrium around the asymptotic value
which is one of the requirement of the WOSSR test. However, this test also requires 
the independence of samples, which is not the case here because subsequent steps in 
Metropolis random walks are indeed correlated, as has been emphasized. Thus, the results
of WOSSR test in this context should be taken with much care and could be misleading 
if $M \ll T_{auto}$. If a large fluctuation from equilibrium value occurs at some 
point, its persistency in sign is fed by the positive correlations of adjacent points 
and this will drive the test to failure even if equilibrium was actually achieved. 
Conversely, it is quite unlikely that the test yields a positive answer on a 
sufficiently large set of running bins if equilibrium is not achieved, unless two 
accidentally large fluctuations of equal time size arise. Therefore, albeit not 
appropriate in principle, the WOSSR test with $M \sim T_{auto}$ provides a fairly 
good indication of equilibrium. 

In order to actually define an equilibrium point, and thereby a relaxation time
$T_{rel}$, we first calculate the arithmetic mean of the rightmost 50 (which is of 
the same order as of $T_{auto}$) bin contents, which is taken as median $X_m$ to 
be fed in the test. The WOSSR test is then
performed on subsets of $M=10$ running bins in the step hystogram taking as starting bin 
the leftmost and moving rightward by one bin at a time. The test is regarded as
successful if it yields a confidence level of at least 0.05 for a median differing
from the previously defined $X_m$ at most by 0.5\%. If the test is successful for 
10 starting bins in a row, the first of those ten is taken to be the equilibrium
point. Little variation of the equilibrium point is found by changing the number 
of running bins from 5 to 20. 

Henceforth, we will use the observable overall multiplicity and WOSSR test to 
point out some important features of the Metropolis algorithm.

%+++++++++++++++++++++++++++++++++++++++++++++++++++++++++
\subsection{Dependence on the integration method}
%+++++++++++++++++++++++++++++++++++++++++++++++++++++++++ 

\begin{figure}
\resizebox{0.5\textwidth}{!}{
  \includegraphics{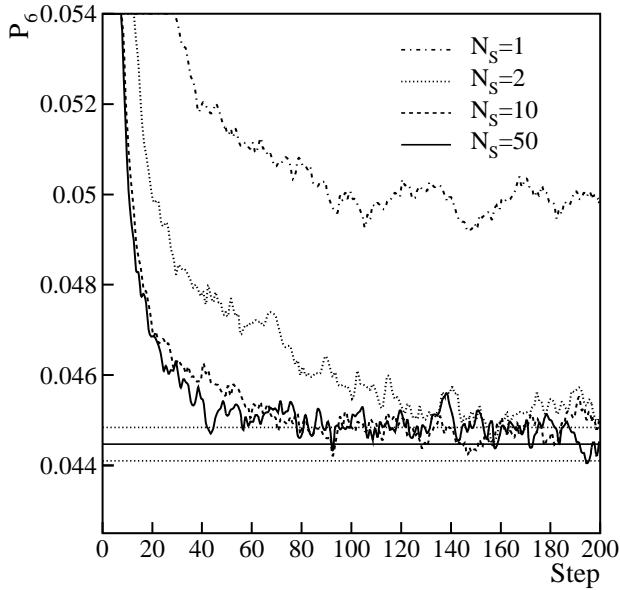}
}  
\caption{Step hystogram showing the convergence to equilibrium of the 
probability $P_{6}$ of a channel with 6 primary hadrons as a function of the 
step in a set of 500000 Metropolis random walks for different numbers of
samples $N_S$ in the Monte-Carlo integration of the phase space volume of each 
channel $\Omega_\Nj$ in Eq. (\ref{Psi}). The horizontal solid line indicates 
the $P_{6}$ value estimated with the importance sampling method and the dashed 
band its relevant statistical uncertainty. Cluster mass is 8 GeV, energy density 
0.4 GeV/fm$^3$ and charges are free.}
\label{step-int}
\end{figure}

It has been shown in Sect.~4 that the importance sampling integration method benefits 
from drawing samples in the extended space of the variables:
\begin{equation}
   \{ N_1, \ldots, N_K | r_1, \ldots, r_{N-1} \}  \qquad {\rm with} \;\; 
   N = \sum_j N_j
\end{equation}
instead of calculating the integral in the $r$'s for each channel $\Nj$ in
Eq. (\ref{Psi}) with a fixed (say 1000) number of Monte-Carlo samples. One could 
try to apply the same idea to the Metropolis algorithm: instead of making a random 
walk in the space of channels $\Nj$ and evaluating the weight of the channel at 
each step by performing a Monte-Carlo integration of $\Omega_\Nj$ in Eq. (\ref{Psi}), 
samples can be drawn in the extended space of the above variables and evaluating 
the integrand function $\Psi$ only one time at each step, thus saving CPU time. 
However, the relaxation time $T_{rel}$ will be different in these two cases and 
most likely longer in the extended space sampling case. This can be easily 
understood if the single $\Psi$ evaluation is regarded as the extreme approximation 
of the integral in the $r$'s in Eq. (\ref{Psi}) with $N_S=1$. This is shown indeed in 
fig. \ref{step-int}: the convergence to the equilibrium speeds up if the number of 
samples used to estimate the integral in Eq. (\ref{Psi}) increases. A serious 
drawback of the extended space sampling, i.e. with $N_S=1$, 
is that the relaxation time may become so long that many bins of the multiplicity 
distribution look like having reached their stable asymptotic value even when 
they actually still slowly drift towards it. We have checked this for the case shown 
in fig. \ref{step-int} by pushing the Metropolis random walk to 10000 steps, much 
beyond the scale of the step hystogram. It has been found 
that even at such large number of steps, the asymptotic value, calculated independently
with the importance sampling method, is not attained, though the WOSSR test yields 
a positive response, under most circumstances. This finding indicates that the use 
of Metropolis algorithm requires more care than expected. At least a comparison 
between the apparent asymptotic stable values with different number of samplings 
or a cross-check with independent calculations is necessary. 

%---------------------------------------------------------------------
%---------------------------------------------------------------------
 
%+++++++++++++++++++++++++++++++++++++++++++++++++++++++++
\subsection{Dependence on the updating rule}
%+++++++++++++++++++++++++++++++++++++++++++++++++++++++++ 

In order to show the effectiveness of the updating rule based on randomly 
sampling the MPD, as discussed in Sect.~4, we have compared it with a simpler 
updating rule based on milder changes of the current configuration. This rule 
is as follows:
\begin{enumerate}
\item{} Three probabilities $\eta_0$, $\eta_+$ and $\eta_-$ are chosen such that 
$\eta_0+\eta_++\eta_-=1$. 
\item{} A random extraction $(0,+,-)$ is made according to the probability 
distribution defined by the $\eta$'s.
\item{} Depending on whether the outcome is $0,+,-$ the overall number of particles 
in the configuration is kept, is increased by 1 unit or decreased by 1 unit 
respectively. In the first case, a randomly chosen particle in the current 
configuration is replaced with one having a mass just above or below. In the 
second case, a new particle is randomly chosen among all possible species. In the 
third case, a randomly chosen particle of the current configuration is removed.
\end{enumerate}
For this updating rule, the proposal matrix $T(m \rightarrow n)$ has been determined
and optimal acceptance matrix set accordingly (see Eq. (\ref{accept})).  

\begin{figure}
\resizebox{0.5\textwidth}{!}{
  \includegraphics{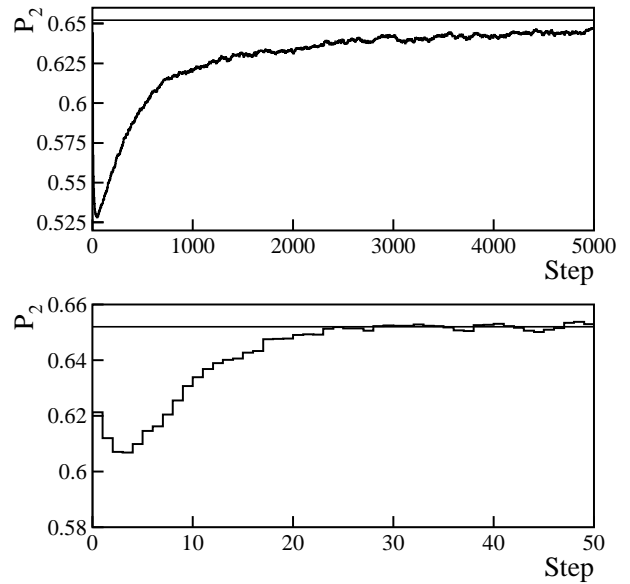}
}  
\caption{Step hystogram showing the convergence to equilibrium of the
probability $P_{2}$ of channels with 2 hadrons  as a function of the step in 
a set of 100000 Metropolis random walks for the updating rule based on 
the MPD (bottom) and on a simpler method described in the text (top). 
Solid lines indicate the true value obtained by computing the phase space
volume of all of the channels. The cluster mass is 2 GeV, the energy density 
0.4 GeV/fm$^3$ and charges are free. The number of samples used in the numerical 
integration of Eq. (\ref{Psi}) was $N_S=1000$. For this comparison, resonances have
been kept at a fixed mass and quantum statistics terms in $\Omj$ have
been neglected.}
\label{step-update}
\end{figure}

Indeed, this rule involves a slowing down of the convergence to equilibrium 
because the fraction of rejected transitions is much higher than in the MPD-based
updating rule. This is apparent in fig. \ref{step-update} where the relevant step 
hystograms are shown for the multiplicity distribution bin with 2 particles for a 
2 GeV mass cluster. Whilst in the MPD-based updating rule 
the equilibrium is achieved within few tens of steps, in the above rule stability is 
not achieved even after 3000 steps. Similar differences are found for heavier clusters. 
Therefore, the MPD-based updating rule is much more efficient with regard to 
computing time.  

%---------------------------------------------------------------------
%---------------------------------------------------------------------

%+++++++++++++++++++++++++++++++++++++++++++++++++++++++++
\subsection{Dependence on the cluster mass and charge}
%+++++++++++++++++++++++++++++++++++++++++++++++++++++++++ 

\begin{figure}
\resizebox{0.5\textwidth}{!}{
  \includegraphics{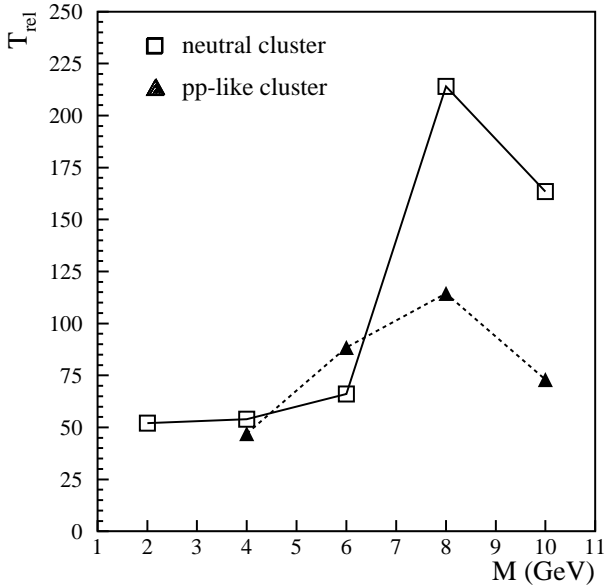}
}  
\caption{Relaxation times in a Metropolis random walk (see text for the 
definition) as a function of mass and charges of the cluster. Energy
density is kept fixed at 0.4 GeV/fm$^3$. Lines are drawn to guide the eye.}
\label{relax}
\end{figure}

We have studied the dependence of $T_{rel}$ on the cluster mass and charges, 
at constant energy density, by applying the WOSSR tests to step hystograms 
of multiplicity distribution $P_n$ in different bins. For each mass and set 
of charges we have taken the highest $T_{rel}$ among the bins for which $P_n
 > 10^{-3}$, as $the$ relaxation time for that cluster. This study 
has been carried out for completely neutral and pp-like clusters; energy density 
has been kept constant at 0.4 GeV/fm$^3$. For each cluster 500000 Metropolis 
random walks (100000 for $M=10$ GeV) have been performed up to 300 steps. 
The relaxation times are shown in fig. \ref{relax}. 

%---------------------------------------------------------------------
%---------------------------------------------------------------------

This just defined relaxation time shows an initial increase as a function
of the cluster mass and drops thereafter going from 8 to 10 GeV. For the present, 
we do not have a complete understanding of this behaviour. An increase as a 
function of the mass is expected because the number of channels sampled from 
the MPD whose weight is much larger than their correspoding microcanonical one 
increases owing to the persistence of the shape difference between multiplicity 
distributions in the canonical and microcanonical ensemble 
(see figs. \ref{eedist},\ref{ppdist}), as already mentioned in Sect.~5. 
The drop, as well as the observed difference between neutral and pp-like cluster, 
in the region 8-10 GeV might be a genuine effect due to a local minimal distance 
between MPD and microcanonical distributions or an artefact of the cut on 
probability at $10^{-3}$. What it is important to remark is that the relaxation
times are not larger than ${\cal O}(100)$ in the region where microcanonical
calculations are necessary. In order to further improve Metropolis calculations, 
the same modification of the sampling distribution put forward at the end of 
Sect.~4 to reduce statistical error in the importance sampling, could be carried 
over here so as to speed up the convergence to equilibrium.

%+++++++++++++++++++++++++++++++++++++++++++++++++++++++++
\subsection{Comparison with the importance sampling method}
%+++++++++++++++++++++++++++++++++++++++++++++++++++++++++ 

The performances of importance sampling method and Metropolis algorithm have
been compared by studying the statistical error on the calculation of averages of 
several observables with the same computing resources. For a neutral cluster with 
4 GeV mass and 0.4 GeV/fm$^3$ energy density, we have calculated the total average 
multiplicity and the primary multiplicity distribution 100 times, each time taking 
100000 steps in both methods. As the sampled distribution at each step is the MPD 
in both cases, the used CPU time is approximately the same. The function $\Psi$ in 
Eq. (\ref{Psi}) has been sampled one time per channel in both methods.

\begin{figure}
\resizebox{0.5\textwidth}{!}{
  \includegraphics{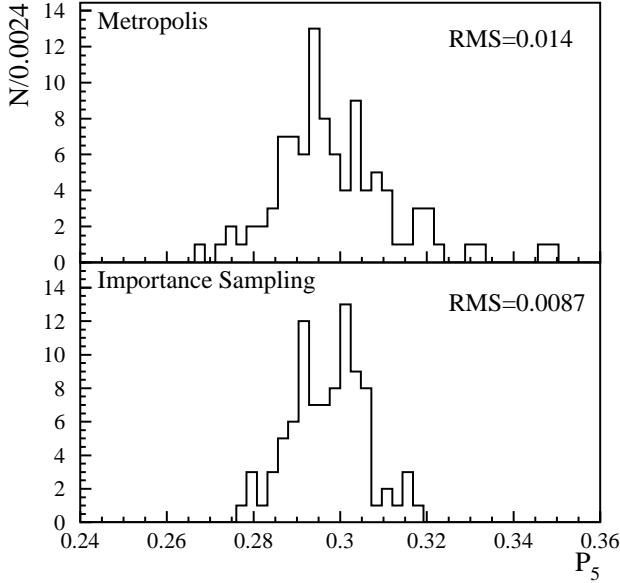}
}  
\caption{Statistical distribution of the Monte-Carlo estimates of the 
probability $P_5$ of channels with 5 primary hadrons, for a neutral cluster with 
4 GeV mass and 0.4 GeV/fm$^3$ energy density.}
\label{compa}
\end{figure}

As an example, we show in fig. \ref{compa} the results obtained for the estimate 
of the probability $P_5$ of channels with 5 primary hadrons. It can be seen 
that the Metropolis algorithm gives rise to a broader statistical distribution of
the estimated values with respect to the importance sampling method. Moreover, the 
distribution is not gaussian and it is slightly asymmetric with also few cases of 
outranging estimates. On the other hand, the 
distributions for the total average multiplicity look quite similar in the two 
methods. It is worth pointing out that the {\em a priori} statistical error estimate 
obtained by using Eq. (\ref{error}) for the importance sampling method is about 
0.008, in good agreement with the found RMS of 0.0087 quoted in fig. \ref{compa}.

We have not investigated in much detail the sources of such a different statistical 
resolution, but we surmise that the ultimate reason of a larger error in Metropolis 
algorithm is the extra call to the random number generator which, at each step, is 
possibly needed to accept or reject a proposed transition. Ultimately, this is 
an additional source of fluctuations in the Metropolis random walk which is absent 
in the importance sampling method where each extracted configuration is simply 
reweighted. Altogether, the importance sampling method seems to be better performing 
to calculate averages in the microcanonical ensemble.

%---------------------------------------------------------------------
%---------------------------------------------------------------------

%***********************************************************************
\section{Conclusions}
%***********************************************************************    
 
We have calculated averages in the microcanonical ensemble of the ideal 
hadron-resonance gas including all light-flavoured known resonances up to 
a mass of about 1.8 GeV. We have found that microcanonical average multiplicities 
of different hadron species differ by less than 10\% from the corresponding 
canonical average (i.e. calculated by introducing a temperature) for clusters 
with relatively low mass, around 8 GeV, and energy density of 0.4 GeV/fm$^3$. 
This confirms and extends previous findings \cite{liu} obtained with a restricted 
sample of hadron species. However, multiplicity distributions in the two ensembles 
show a clear difference in shape which seem to persist in the thermodynamical limit.
Particularly, the dispersion of the total multiplicity distribution is, to
a great accuracy, 1/2 of the Poisson dispersion in the large volume limit; 
presently, we do not have any simple explanation of this fact and whether this
is linked to our special choice of the energy density.

A major point of this study is concerned with the numerical methods developed 
to calculate microcanonical averages: an importance sampling method, which has 
been proposed and used in this work for the first time; and a previously 
proposed Metropolis algorithm \cite{weai}, whose performances have been improved 
by using as proposal matrix Poisson distributions with mean values set as those 
of the grand-canonical ensemble. They have been also employed as sampled 
distributions in the importance sampling method. 

The Metropolis algorithm is capable to provide single samples of the microcanonical 
ensemble with unit weight and it is thus a suitable tool for event generation. 
Yet, it was proved to be less accurate than the importance sampling method for the 
calculation of averages for single clusters. Moreover, the Metropolis algorithm 
used for event generation requires much care, particularly a preliminary study of 
how fast the equilibrium condition is achieved. The convergence to equilibrium 
may depend on the observable to be analyzed and, more worrying, on the specific 
integration methods to evaluate the microcanonical weights of the channels.

However, the present study indicates that there is still room for a further 
improvement of the efficiency of both examined methods. An efficient way of
calculating microcanonical ensemble opens the way to test the statistical 
hadronization model at low energy and with respect to many more observables than
those considered as yet.

%**********************************************************************
\section*{Acknowledgements}
%**********************************************************************

We are grateful to J. Aichelin, T. Gabbriellini, M. Gorenstein, A. Ker\"anen,
K. Werner for useful discussions. This work has been carried out within the 
INFN research project FI31.

%******************************************************************************
\section*{Appendix}    
%******************************************************************************    
\appendix
\renewcommand{\theequation}{\thesection.\arabic{equation}}

%******************************************************************************
\setcounter{equation}{0}
\section{Calculation of the phase space integrals}    
%******************************************************************************    

Here we summarize a method to calculate phase space integrals due to Hagedorn 
\cite{cerhag1}. 
An analytical calculation of $\Phi$ in Eq. (\ref{Phi}) can be carried out for two
or three particles whilst for more particles the expressions get rapidly 
so complicated that a numerical computation is much more suitable.

For two particles:
\begin{equation}\label{Phi2}
\Phi(M,m_1,m_2)= \frac{4 \pi \p^*}{T^2} \frac{\epsilon_1 \epsilon_2}{M}
\end{equation}
where $m_i$'s are the masses, $\epsilon_i = \sqrt{\p^{*2}+m_i^2}$ the energies,
$M$ the cluster's mass, $T$ the total available kinetic energy and: 
\begin{equation}\label{pistar}
\p^*=\frac{1}{2}\bigg[ M^2-2(m_1^2+m_2^2)+
\frac{1}{M^2}(m_1^2-m_2^2)^2 \bigg]^{\frac{1}{2}}. 
\end{equation}
For $N>2$ a function $W$ of the momenta $\p_1,\ldots,\p_N$ is introduced:
\begin{equation}\label{phiw}
\Phi = \frac{(4\pi)^N}{T^{3N-4}} \int \Big[ \prod_{i=1}^{N} \d{\p}_i \; 
 {\p}_i^2 \Big] \, \delta(M-\sum_{i=1}^{N}{\epsilon}_i)\,W(\p_1,\ldots,\p_N)
\end{equation}
The function $W$ reads:
\begin{equation}\label{wfun}
 W(\p_1,\ldots,\p_N)=\frac{1}{(4\pi)^N} \int \biggl[ \prod_{i=1}^{N} 
 \d \Omega_i \biggr] \;\delta^3(\sum_{i=1}^{N} \p_i \hat{\bf p}_i) 
\end{equation} 
where $\hat {\bf p}_i = {\bf p}_i/\p_i$, and can be calculated explicitely:
\begin{eqnarray}\label{wfun2}
\!\!\!\!\!\!\!\!\!\! 
W(\p_1,\ldots,\p_N) &=& - \frac{1}{2^{N+1}\pi(N-3)! \p_1 \ldots \p_N} \nonumber \\
\!\!\!\!\!\!\!\!\!\! &\times& \!\!\!\!\! 
 \sum_{\begin{array}{ll}{\scriptstyle \{ \sigma_1 \ldots \sigma_N\}}
 \\{\scriptstyle \sum \sigma_j \p_j \geq 0} \end{array}}
 \sigma_1 \ldots \sigma_N \big( \sum_i \sigma_i \p_i \big)^{N-3}
\end{eqnarray}  
where $\sigma$ can be either $+1$ or $-1$. This expansion involves a large number 
of terms even at relatively small $N$, so it is more advantegeous under some 
circumstances, to calculate $W$ differently. In fact, one can set in Eq. (\ref{wfun}):
\begin{equation}
 \delta^3(\sum_{i=1}^N \p_i \hat{\bf p}_i) = \frac{1}{(2\pi)^3} \int \d^3 {\u} 
  \; \exp\big[ -\i \sum_{i=1}^N \p_i \hat{\bf p}_i \cdot {\bf u} \big]
\end{equation}
which leads to \cite{weai}:
\begin{equation}\label{wfun3}
 W(\p_1,\ldots,\p_N)= \frac{1}{2\pi^2} \int_0^\infty \d \u \; \u^2 \prod_{i=1}^N
 \frac{\sin (\p_i \u)}{\p_i \u}
\end{equation} 
Setting now $y=\u \mu$, where $\mu$ is an arbitrary energy scale, to make 
integration variable adimensional:
\begin{eqnarray}\label{wfun4}
\!\!\!\!\!\!\!\! && W(\p_1,\ldots,\p_N) = \frac{1}{2\pi^2 \mu^3} \int_0^\infty \d y 
 \; y^2 \prod_{i=1}^N \frac{\sin [(\p_i/\mu) y]}{(\p_i/\mu) y} \nonumber \\
\!\!\!\!\!\!\!\! && = \frac{1}{2\pi^2 \mu^{3}} \int_0^1 \d x \; 
 \frac{(1-x)^{N-4}}{x^{N-2}}
 \prod_{i=1}^N \frac{\sin \left( \frac{\p_i}{\mu} \frac{x}{1-x} \right)}{p_i/\mu}
\end{eqnarray} 
where the last expression has been obtained by using the variable transformation
$y=x/(1-x)$.
The last integral can be calculated numerically provided that the number of 
particles is at least of the order of 10, otherwise the integrand function 
features strong oscillations which make most numerical integration methods 
failing. In this case, it is compelling to calculate $W$ by means of its full 
expansion (\ref{wfun2}). However, the CPU time needed to sum all of the terms 
rapidly grows with $N$, so that for $N > 10$ it is definitely preferrable 
to switch to the numerical computation of the integral~(\ref{wfun4}) (see
also fig. \ref{timechann}). The arbitrary scale $\mu$ can be set so as to 
obtain a good resolution in the numerical computation of $W$ and we have found
that $\mu = 1$ GeV is an appropriate choice for most cases at the energy
densities we have been dealing with; if the number of
momenta larger than 100 MeV is $\le 10$, we set $\mu = 100$ MeV. 

\begin{figure}
\resizebox{0.5\textwidth}{!}{%
  \includegraphics{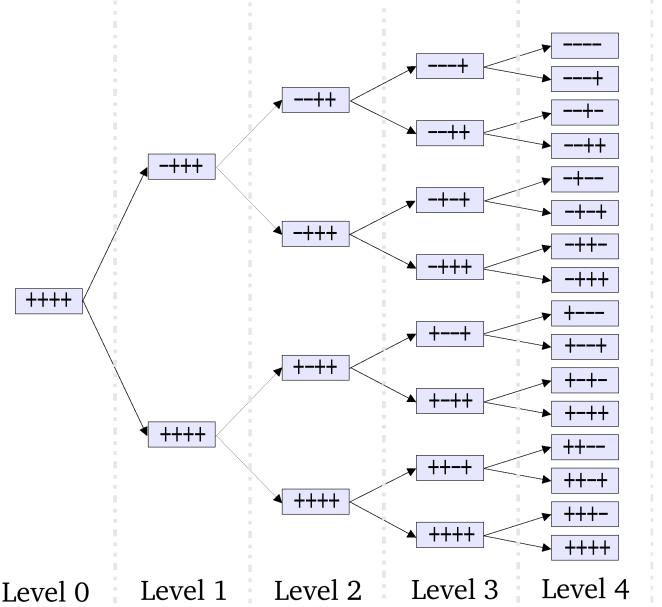}
}
\caption{Tree diagram to calculate $W$ with (\ref{wfun2}) for 4 particles.}
\label{tree}
\end{figure}

In order to minimize the CPU time spent to calculate $W$ by using its full 
expansion (\ref{wfun2}) we have optimized the loop over all combinations 
$\{\sigma_1,\ldots,\sigma_N\}$ such that $\sum \sigma_j \p_j > 0$ by considering 
only those which are relevant. 
The momenta  $\p_1,\ldots,\p_N$ are first sorted such that $\p_1> \p_2>\ldots> \p_H$
and all possible $N$-uples made of $+1$ or $-1$ are arranged in a multi-level tree 
structure shown in fig. \ref{tree} for the special case $N=4$. 
%---------------------------------------------------------------------------------
%----------------------------------------------------------------------------------
If, for some $N$-uple at some level $\sum_{j=1}^{H} \sigma_j \p_j \leq 0$, 
the same is true for all $N$-uples branching from it because of momentum 
sorting and of the tree structure, which is such that moving upwards one level a 
change in sign $(+ \rightarrow -)$ occurs. Note that replica $N$-uples in 
fig. \ref{tree} are just meant to show the tree structure, though they are in fact 
not considered in the actual loop.  
 
The phase space integral in Eq. (\ref{phiw}) is now transformed by means of 
a sequence of changes of integration variable \cite{weai} $p_i \rightarrow t_i
\rightarrow s_i \rightarrow x_i \rightarrow z_i \rightarrow r_i$. If $t_i$ denote
the particle kinetic energies and $T$ the total available kinetic energy:
\begin{eqnarray}\label{transf}
&& p_i=\sqrt{t_i(t_i+2m_i)} \qquad i=1,\ldots,N  \nonumber \\
&& t_i=s_i-s_{i-1} \qquad i=1,\ldots,N \; \textrm{with $s_0=0$ and $s_n = T$} 
\nonumber \\
&& x_i=\frac{s_i}{T} \qquad i=1,\ldots,N-1 \nonumber \\
&& z_i=\frac{x_i}{x_{i+1}} \qquad i=1,\ldots,N-1 \;
\textrm{with $x_N\equiv1$} \nonumber \\
&& r_i= z_i^i \qquad i=1,\ldots,N-1
\end{eqnarray}
As a result of this sequence of transformations, the Dirac delta of energy 
conservation in Eq. (\ref{phiw}) is integrated away and the phase space integral
reads:
\begin{equation}\label{phiupsi}
 \Phi = \int_0^1 \d r_1 \ldots \int_0^1 \d r_{N-1} \Upsilon (r_1,\ldots,r_{N-1})
\end{equation}
where:
\begin{equation}
 \Upsilon(r_1,\ldots,r_{N-1}) = \frac{(4\pi)^N T^{3-2N}}{(N-1)!}
 \prod_{i=1}^N p_i \epsilon_i W(p_1,\ldots,p_N)
\end{equation}
and $p_1,\ldots,p_N$ are to be calculated by going along the transformations~
(\ref{transf}) for a given set $(r_1,\ldots,r_N)$. In the form (\ref{phiupsi})
$\Phi$ can be easily calculated by means of Monte Carlo integration.

%******************************************************************************
\setcounter{equation}{0}
\section{Error estimates for importance sampling}    
%******************************************************************************    

We have defined the best estimate of the mean value of the observable $O$ in 
Eq. (\ref{estimator}). This estimator can be viewed as a random variable because 
the channels $\Nj$ are random variables whose distribution is $\Pi$ in 
Eq. (\ref{poisson}). Let us introduce the random variables $\sf \Omega$, $\sf O$ 
and $\sf \Pi$ taking values $\Omj$, $O(\Nj)$ and $\Pij$ respectively for a 
particular channel $\Nj$. The following relations involving expectation values
hold:
\begin{eqnarray}\label{expect}
\!\!\!\!\!\!\!\!\!\! && {\sf E}_\Pi \left( {\displaystyle \frac{\sf \Omega}{\sf \Pi}} 
 \right) = \sum_{\Nj} {\displaystyle \frac{\Omj}{\Pij}} \Pij = \Omega \nonumber \\
\!\!\!\!\!\!\!\!\!\! && {\sf E}_\Pi \left( {\sf O} {\displaystyle \frac{\sf \Omega}
 {\sf \Pi}} \right) = \sum_{\Nj} O(\Nj) {\displaystyle \frac{\Omj}{\Pij}} \Pij = \Omega 
 \langle O \rangle
\end{eqnarray}
where Eqs.~(\ref{microdec}) and (\ref{obs}) have been used.
We can now write the estimator random variable $\langle {\sf O} \rangle$ from
Eq. (\ref{estimator}) as:
\begin{equation}\label{ranvar}
  \langle {\sf O} \rangle = \frac{\sum_{k=1}^M {\sf O}_k 
  {\displaystyle \frac{{\sf \Omega}_k}{{\sf \Pi}_k} }}
  {\sum_{k=1}^M {\displaystyle \frac{{\sf \Omega}_k}{{\sf \Pi}_k}}}
\end{equation}
If the number of events $M$ is large enough the numerator and denominator in the
right hand side are gaussianly distributed according to the central limit theorem
and the variance $\sigma^2(\langle{\sf O}\rangle)$ can be calculated by using the 
error propagation formula:
\begin{equation}\label{variance0}
  \sigma^2(\langle {\sf O} \rangle ) = \sigma^2({\sf N})\frac{1}{D^2} +
  \sigma^2({\sf D}) \frac{N^2}{D^4} -2 {\rm cov}({\sf N,D}) \frac{N}{D^3}
\end{equation}
where ${\sf N}$ and ${\sf D}$ are the numerator and the denominator in 
Eq. (\ref{ranvar}), and $N$ and $D$ their expectation values ${\sf E}_\Pi({\sf N})$ 
and ${\sf E}_\Pi({\sf D})$ respectively. Because of Eq. (\ref{expect}):
\begin{equation}\label{expect2}
 N \equiv  {\sf E}_\Pi({\sf N}) = M \Omega \langle O \rangle \qquad 
 D \equiv  {\sf E}_\Pi({\sf D}) = M \Omega 
\end{equation}
Thus, by using Eq. (\ref{expect2}) and the general variance definitions:
\begin{eqnarray}
 \sigma({\sf x}) &=& {\sf E}({\sf x}^2)- {\sf E}({\sf x})^2 \nonumber \\
{\rm cov}({\sf x},{\sf y}) &=&  {\sf E}({\sf xy})- {\sf E}({\sf x})
{\sf E}({\sf y})
\end{eqnarray}
we obtain:
\begin{eqnarray}\label{covaria}
\sigma^2({\sf N}) \frac{1}{D^2} &=&  \frac{1}{M \Omega^2} \left[ {\sf E}_\Pi \left( 
 {\sf O}^2{\displaystyle \frac{{\sf \Omega}^2}{{\sf \Pi}^2}} \right) - 
 \langle O \rangle^2 \Omega^2 \right] \nonumber \\
\sigma^2({\sf D}) \frac{N^2}{D^4} &=& \frac{\langle O \rangle^2}{M \Omega^2} 
\left[ {\sf E}_\Pi \left({\displaystyle \frac{{\sf \Omega}^2}{{\sf \Pi}^2}}\right) 
 - \Omega^2 \right] \nonumber \\
{\rm cov}({\sf N,D}) \frac{N}{D^3} &=& \frac{\langle O \rangle}{M \Omega^2}
\left[ {\sf E}_\Pi \left({\sf O} {\displaystyle \frac{{\sf \Omega}^2}{{\sf \Pi}^2}}
\right) - \langle O \rangle \Omega^2 \right] 
\end{eqnarray}
Therefore Eq. (\ref{variance0}) can be rewritten as:
\begin{eqnarray}\label{variance}
\sigma^2(\langle {\sf O} \rangle) &=&  \frac{1}{M \Omega^2} \Bigg\{ \left[ 
{\sf E}_\Pi \left( {\sf O}^2 {\displaystyle \frac{{\sf \Omega}^2}{{\sf \Pi}^2}}  
 \right) - \langle O \rangle^2 \Omega^2 \right] \nonumber \\
&+& \langle O \rangle^2 \left[ {\sf E}_\Pi \left( 
{\displaystyle \frac{{\sf \Omega}^2}{{\sf \Pi}^2}} \right)- \Omega^2 \right] \nonumber \\ 
&-& 2 \langle O \rangle \left[ {\sf E}_\Pi \left( {\sf O} 
 {\displaystyle \frac{{\sf \Omega}^2}{{\sf \Pi}^2}} \right) - \langle O \rangle 
 \Omega^2 \right] \Bigg\} 
\end{eqnarray}
which is essentially the Eq. (\ref{error}). 

The estimator (\ref{estimator}) is a biased one. This can be proved writing 
Eq. (\ref{ranvar}) as $\langle {\sf O} \rangle = {\sf N}/{\sf D}$ and: 
\begin{eqnarray}\label{bias0}
&& {\sf E}_\Pi \left(\frac{\sf N}{\sf D}\right) = {\sf E}_\Pi ({\sf N})
 {\sf E}_\Pi \left(\frac{1}{\sf D}\right) + {\rm cov}({\sf N},1/{\sf D}) \nonumber \\
&& \simeq {\sf E}_\Pi ({\sf N}) \left[ \frac{1}{{\sf E}_\Pi ({\sf D})} +
 \frac{\sigma^2({\sf D})}{{\sf E}_\Pi ({\sf D})^3} \right] - 
 \frac{{\rm cov}({\sf N},{\sf D})}{{\sf E}_\Pi ({\sf D})^2}
\end{eqnarray}
using the definition of covariance and expanding the function 1/D around its
mean value. The bias $B$ of the estimator (\ref{ranvar}) can then be written as:
\begin{eqnarray}\label{bias}
B &\equiv& {\sf E}_\Pi (\langle {\sf O} \rangle) - \langle O \rangle = 
 {\sf E}_\Pi \left(\frac{\sf N}{\sf D}\right) - \langle O \rangle \nonumber \\
 &\simeq& \frac{1}{M \Omega^2} \left[ \langle O \rangle {\sf E}_\Pi \left( 
{\displaystyle \frac{{\sf \Omega}^2}{{\sf \Pi}^2}} \right) - 
 {\sf E}_\Pi \left({\sf O} {\displaystyle \frac{{\sf \Omega}^2}{{\sf \Pi}^2}}\right) 
 \right] 
\end{eqnarray} 
by using Eqs. (\ref{bias0}),(\ref{covaria}),(\ref{expect2}).

 Estimates of the bias (\ref{bias}) and the variance (\ref{variance}) can be 
obtained by replacing in Eq. (\ref{variance}) the expectation values with arithmetic 
means:
\begin{equation}
   {\sf E}_\Pi \rightarrow \frac{1}{M} \sum_{k=1}^M
\end{equation}
while $\langle O \rangle$ can be estimated through Eq. (\ref{estimator}) and
the best estimate of $\Omega$ is, according to Eq. (\ref{expect}):
\begin{equation}
  \Omega = {\sf E}_\Pi \left( {\displaystyle \frac{\sf \Omega}{\sf \Pi}} \right) 
  \doteq \frac{1}{M} \sum_{k=1}^M {\displaystyle \frac{\Omj^{(k)}}{\Pij^{(k)}}}
\end{equation}
The estimator (\ref{estimator}) could be corrected for the bias by subtracting the 
estimate of (\ref{bias}). However, the bias (\ref{bias}) is proportional to $1/M$ 
unlike the statistical error $\sigma_{\langle O \rangle}$ decreasing like 
$1/\sqrt{M}$ and, thus, becomes negligible with respect to the statistical error for 
a large number of samplings.
 
%******************************************************************************
\setcounter{equation}{0}
\section{Multiplicity distribution in the canonical ensemble}    
%******************************************************************************    

The multi-species multiplicity distribution in the canonical ensemble with 
internal abelian charges $\Qz$, can be obtained from the generating function 
associated to the canonical partition function $Z(\Qz)$. The probability of a 
single state in the canonical ensemble reads:
\begin{equation}\label{prob}
  P_{\rm state} = \frac{1}{Z(\Qz)} \; \e^{-E_{\rm state}/T} \delta_{\Qz_{\rm state},\Qz}
\end{equation}
and the generating function:
\begin{equation}\label{gener1}
 G(\lambda_1,\ldots,\lambda_K) = \frac{1}{Z(\Qz)}\sum_{\rm states} 
 \e^{-E_{\rm state}/T} \delta_{\Qz_{\rm state},\Qz} \prod_{j=1}^K 
 \lambda_j^{N_{j\,{\rm state}}}
\end{equation}
where $K$ is the number of hadron species and $N_{j\,{\rm state}}$ the number of
hadrons of species $j$ in the state. The generating function can be worked out by 
Fourier expanding the Kronecker delta in Eq. (\ref{prob}) \cite{becaheinz,becagp}:
\begin{eqnarray}\label{gener2}
 && G(\lambda_1,\ldots,\lambda_K) = \frac{1}{Z(\Qz)}\frac{1}{(2\pi)^M} 
 \int_{-\pivs}^{\pivs} \d^M \phi \; \e^{\i \Qz \cdot \phivs} \exp \bigg[ \nonumber \\
 && \sum_{j=1}^K \frac{(2J_j+1)V}{(2\pi)^3}
 \int \d^3 \p \; \log (1\pm\lambda_j\e^{-\sqrt{\p^2+m_j^2}/T-\i\qj\cdot\phivs})
 ^{\pm 1} \bigg] \nonumber \\
\end{eqnarray}
where the upper sign applies to fermions, the lower to bosons.

According to Eqs.~(\ref{prob}) and (\ref{gener1}), the probability of a given 
n-tuple of hadrons $\Nj$, that is the multi-species multiplicity distribution, can 
be obtained by integrating the generating function over the unitary circle in the 
complex $\lambda_j$ planes:
\begin{equation}\label{prob2}
 P(\Nj) = \bigg[ \prod_{j=1}^K \frac{1}{2\pi\i} \oint \frac{\d \lambda_j}
 {\lambda_j^{N_j}} \bigg] G(\lambda_1,\ldots,\lambda_K)
\end{equation}
Plugging Eq. (\ref{gener2}) in the (\ref{prob2}) and using the residual theorem,
one obtains:
\begin{eqnarray}\label{prob3}
 && P(\Nj) = \frac{1}{Z}\frac{1}{(2\pi)^M} \int_{-\pivs}^{\pivs} \d^M \phi \; 
 \e^{\i \Qz \cdot \phivs} \nonumber \\
 && \prod_{j=1}^K \frac{1}{N_j!} \frac{\partial^{N_j}}{\partial \lambda_j^{N_j}}
 \exp\bigg[ \sum_{n_j=1}^\infty (\mp 1)^{n_j+1}\frac{z_{j(n_j)}\lambda_j^{n_j}}
 {n_j}\e^{-n_j\i\qj\cdot\phivs} \bigg] \Bigg|_{\lambda=0} \nonumber \\
 &&
\end{eqnarray}
where $z_{j(n)}$ is as in Eq. (\ref{zetap}).
Now the exponential in Eq. (\ref{prob3}) is expanded:
\begin{eqnarray}\label{expand}
 &&\exp\left[ \sum_{n_j=1}^\infty (\mp 1)^{n_j+1}\frac{z_{j(n_j)}\lambda_j^{n_j}}
 {n_j} \; \e^{-n_j\i\qj\cdot\phivs} \right] \nonumber \\
 &=& \prod_{n_j=1}^\infty  \exp\left[ (\mp 1)^{n_j+1}\frac{z_{j(n_j)}
 \lambda_j^{n_j}}{n_j}\;\e^{-n_j\i\qj\cdot\phivs} \right] \nonumber \\
 &=& \prod_{n_j=1}^\infty \sum_{h_{n_j}=0}^\infty (\mp 1)^{(n_j+1)h_{n_j}} 
 \frac{z_{j(n_j)}^{h_{n_j}}\lambda_j^{n_j{h_{n_j}}}}{n_j^{h_{n_j}} h_{n_j}!}\;
 \e^{-n_j h_{n_j}\i\qj\cdot\phivs} \nonumber \\
 &=& \sum_{\substack{h_{n_1}=0 \\ \vdots \\ h_{n_K}=0}}
 ^\infty \prod_{n_j=1}^\infty (\mp 1)^{(n_j+1)h_{n_j}}
 \frac{z_{j(n_j)}^{h_{n_j}}\lambda_j^{n_j{h_{n_j}}}}{n_j^{h_{n_j}} h_{n_j}!}
 \; \e^{-n_j h_{n_j} \i\qj\cdot\phivs}
 \nonumber \\ && 
\end{eqnarray} 
Taking the derivatives of the latter expression with respect to the $\lambda_j$'s 
in $\lambda_j=0$ according to Eq. (\ref{prob3}), one is left with non vanishing 
terms in the above equation if $\sum_{n_j=1}^\infty n_j h_{n_j} = N_j$. 
Therefore, the last series in Eq. (\ref{expand}) gives rise to:
\begin{equation}\label{expand2}
 N_j! \, \e^{-N_j\i\qj\cdot\phivs}
 \sum_{\hpartj}(\mp 1)^{N_j + \sum_{n_j} h_{n_j}} \prod_{n_j}
 \frac{z_{j(n_j)}^{h_{n_j}}}{n_j^{h_{n_j}} h_{n_j}!}  
\end{equation}
where $\hpartj$ indicates the set of {\em partitions} (in the multiplicity representation)
of the integers $N_j$, i.e. integers such that $\sum_{n_j=1}^\infty n_j h_{n_j} = N_j$.
The $n_j$ and $h_j$ indices actually run from 1 to $N_j$. Defining $\sum_{n_j} h_{n_j}
= H_j$ and restoring the Kronecker delta, one can rewrite Eq. (\ref{prob3}) by 
using the expression of derivatives in Eq. (\ref{expand2}) as:
\begin{eqnarray}
\!\!\!\!\!\!\! && P(\Nj) = \frac{1}{Z(\Qz)} \nonumber \\
\!\!\!\!\!\!\! && \times \Bigg[ \prod_{j=1}^K \sum_{\hpartj} 
 (\mp 1)^{N_j + H_j} \prod_{n_j} \frac{z_{j(n_j)}^{h_{n_j}}}
 {n_j^{h_{n_j}} h_{n_j}!} \Bigg] \delta_{\Qz,\sum_j N_j \qj}
\end{eqnarray}
which is Eq. (\ref{candis}).

%******************************************************************************
\setcounter{equation}{0}
\section{Error estimates for Metropolis algorithm}    
%******************************************************************************    

An estimator of the mean value of the observable $O$ in a $M$-steps Metropolis
random walk has been defined in Eq. (\ref{metroestim}). As well as for the importance
sampling method, this estimator can be viewed as a random variable $\langle {\sf O}
\rangle$ and so the values of the observables at each step ${\sf O}^{(k)}$. 
Therefore:
\begin{equation}\label{varia1}
  \sigma^2(\langle {\sf O} \rangle) = {\sf E}(\langle {\sf O} \rangle^2)-
  \langle O \rangle^2 = {\sf E}\left[ \left( \frac{\sum_{k=1}^M {\sf O}^{(k)}}{M}
  \right)^2 \right] - \langle O \rangle^2
\end{equation}
Thence:
\begin{eqnarray}
 && {\sf E}\left[\left(\frac{\sum_{k=1}^M {\sf O}^{(k)}}{M} \right)^2\right] 
 = \nonumber \\
 && = \frac{1}{M^2} \sum_{k=1}^M {\sf E}({\sf O}^{(k)2}) + 
 \frac{2}{M^2} \sum_{k<i} {\sf E}({\sf O}^{(k)}{\sf O}^{(i)}) \nonumber \\
 && = \frac{1}{M} {\sf E}({\sf O}^2) + 
 \frac{2}{M^2} \sum_{k<i}{\sf E}({\sf O}^{(k)}{\sf O}^{(i)}) \nonumber \\
 && = \frac{1}{M} {\sf E}({\sf O}^2) + 
 \frac{2}{M^2} \sum_{k=1}^M \sum_{l=1}^{M-k} A(l) + \langle O \rangle^2
\end{eqnarray}
where $A$ is the autocorrelation function. 
We can then write:
\begin{eqnarray}\label{metrovar}
 && {\sf E}\left[\left(\frac{\sum_{k=1}^M {\sf O}^{(k)}}{M} \right)^2\right] =
 \nonumber \\
 && = \frac{{\sf E}({\sf O}^2)}{M} + \frac{M-1}{M} \langle O \rangle^2 +
 \frac{2}{M^2} \sum_{k=1}^M \sum_{l=1}^{M-k} A(l)
\end{eqnarray}
If $M \gg T_{auto}$, we can approximate the inner sum in Eq. (\ref{metrovar}) with 
the integral of the autocorrelation function $R$ defined as:
\begin{equation}
  R = \sum_{l=1}^\infty A(l)
\end{equation}
and rewrite an approximate expression for 
Eq. (\ref{metrovar}) as:
\begin{equation}\label{metrovar2}
 {\sf E}\left[\left(\frac{\sum_{k=1}^M {\sf O}^{(k)}}{M} \right)^2\right] 
 \simeq  \frac{{\sf E}({\sf O}^2)}{M} + \frac{M-1}{M}\langle O \rangle^2 
 + 2 \frac{R}{M}
\end{equation}
Therefore, Eq. (\ref{varia1}) becomes:
\begin{equation}
  \sigma^2(\langle {\sf O} \rangle) = 
  \frac{\langle O^2 \rangle - \langle O \rangle^2 + 2R}{M} 
\end{equation}  
which proves Eq. (\ref{metroerro}).

%**********************************************************************

%***********************************************************************

\end{document}